\documentclass[journal]{IEEEtran}

\usepackage[cmex10]{amsmath}
\usepackage{cite}

\usepackage{latexsym}
\usepackage{graphics}
\usepackage{amssymb}
\usepackage{amsthm}
\usepackage{cite}
\usepackage{array}
\usepackage{graphicx}

\usepackage{url}

\newcommand{\fracSum}[1]{{\underset{{#1}}{\sum}}}

\newcommand{\fracSumtwo}[2]{\overset{#2}{\underset{#1}{\sum}}}
\newcommand{\vect}[1]{\mathbf{#1}}

\def\Htran{\mbox{\tiny $\mathrm{H}$}}

\newcommand{\maximize}[1]{{\underset{{#1}}{\mathrm{maximize}}}}

\theoremstyle{plain}

\newtheorem{theorem}{Theorem}
\newtheorem{corollary}{Corollary}
\newtheorem{lemma}{Lemma}

\newtheorem{proposition}{Proposition}

\begin{document}

\title{Deploying Dense Networks for Maximal Energy Efficiency: Small Cells Meet Massive MIMO}

\IEEEoverridecommandlockouts

\author{Emil Bj{\"o}rnson, \emph{Member, IEEE}, Luca Sanguinetti, \emph{Senior Member, IEEE}, \\ and Marios Kountouris, \emph{Senior Member, IEEE}
\thanks{E.~Bj\"ornson is with the Department of Electrical Engineering (ISY), Link\"{o}ping University, Link\"{o}ping, Sweden (emil.bjornson@liu.se). L.~Sanguinetti is with the University of Pisa, Dipartimento di Ingegneria dell'Informazione, Pisa, Italy (luca.sanguinetti@iet.unipi.it) and is also with the Large Systems and Networks Group (LANEAS), CentraleSup\'elec, Universit\'e Paris-Saclay, 3 rue Joliot-Curie,  91192 Gif-sur-Yvette, France. M.~Kountouris is with the Mathematical and Algorithmic Sciences Lab, France Research Center, Huawei Technologies Co. Ltd., France (marios.kountouris@huawei.com).
\newline\indent E.~Bj\"ornson is supported by ELLIIT and an Ingvar Carlsson Award from the Swedish Foundation for Strategic Research. L.~Sanguinetti is funded by the People Programme (Marie Curie Actions) FP7 PIEF-GA-2012-330731 Dense4Green and is also supported by the ERC Starting Grant 305123 MORE. 
\newline \indent Preliminary versions \cite{Bjornson2015icc,Bjornson2015spawc} of this paper were presented at the IEEE International Conference on Communications (ICC) 2015 and at the IEEE Workshop on Signal Processing Advances in Wireless Communications (SPAWC) 2015.}}

\maketitle

\begin{abstract}
How would a cellular network designed for maximal energy efficiency look like? To answer this fundamental question, tools from stochastic geometry are used in this paper to model future cellular networks and obtain a new lower bound on the average uplink spectral efficiency. This enables us to formulate a tractable uplink energy efficiency (EE) maximization problem and solve it analytically  with respect to the density of base stations (BSs), the transmit power levels, the number of BS antennas and users per cell, and the pilot reuse factor. The closed-form expressions obtained from this general EE maximization framework provide valuable insights on the interplay between the optimization variables, hardware characteristics, and propagation environment. Small cells are proved to give high EE, but the EE improvement saturates quickly with the BS density. Interestingly, the maximal EE is achieved by also equipping the BSs with multiple antennas and operate in a ``massive MIMO'' fashion, where the array gain from coherent detection mitigates interference and the multiplexing of many users reduces the energy cost per user.
\end{abstract}

\section{Introduction}

\label{sec:intro}

The biggest challenges for next generation wireless communication systems (5G) are to support the ever-growing demands for higher data rates and to ensure a consistent quality of service (QoS) throughout the entire network \cite{Andrews2014a}.  
To meet these demands, the network area throughput (in bit/s/km$^2$) needs to increase by a {factor of 1000} over the next 10--15 years \cite{Qualcomm}. At the same time, the power consumption of the information and communication technology (ICT) industry and the corresponding energy-related pollution are becoming major societal and economical concerns \cite{Fehske2011a}. Credited sources foresee that, to meet such a $1000 \times$ higher data traffic without increasing the ICT footprint, new technologies that improve the overall energy efficiency (EE) by $1000 \times$ must be developed \cite{Greentouch2013}. Hence, higher network area throughput on the one hand and less power consumption on the other are  seemingly contradictory 5G requirements \cite{Bjornson2014c}. There is a broad consensus that these goals can only be jointly achieved by a substantial network densification.

Two promising technologies towards network densification are small-cell networks \cite{Hoydis2011c,Sanguinetti2015bJSAC} and massive MIMO systems \cite{Marzetta2010a,Larsson2014a,Yang2013b}. The first technology relies on an ultra-dense and irregular operator-deployment of low-cost and low-power base stations (BSs), with higher density where the user load is higher. Bringing the BSs and user equipments (UEs) closer to each other can increase the area throughput, while significantly reducing the radiated signal power. However, this may come at the price of a substantial increase of the circuit power consumption (per km$^2$) due to the larger amount of hardware and infrastructure \cite{Bjornson2013e}. In contrast, the massive MIMO technology aims at evolving the conventional BSs by using arrays with a hundred or more small dipole antennas. This allows for coherent multi-user MIMO transmission where tens of users can be multiplexed in both the uplink (UL) and the downlink (DL) of each cell. It is worth observing that, contrary to what the name ``massive'' suggests, massive MIMO arrays are rather compact; 160 dual-polarized antennas at 3.7 GHz fit into the form factor of a flat-screen television \cite{Vieira2014a}.
In massive MIMO systems, the area throughput is improved by the multiplexing gain, while the array gain from coherent processing allows for major reductions in the emitted power. Similar to small-cell networks, however, the potential throughput gains from massive MIMO come from deploying more hardware (i.e., multiple antenna branches per BS), which in turn increases the circuit power consumption per BS.

In summary, both densification technologies can improve the area throughput and reduce the radiated power, but at the cost of deploying more hardware infrastructure. Hence, the overall EE of the network can only be improved if these benefits and costs are properly balanced. The main objective of this paper is to develop an analytical framework for designing dense cellular networks for maximal EE, as well as to provide guidelines for practical deployment when (among others) the spectral efficiency (SE), radiated power, BS density, number of antennas and UEs per BS, channel estimation, and the circuit power consumption are taken into account from the outset.

\subsection{Related Works}

The EE of cellular networks has been defined from different perspectives in the last decade \cite{Feng2013a,Li2011a}. One of the most common definitions is a benefit-cost ratio, where the service quality per area unit (a.u.) is compared with the associated energy costs. In this paper, the following general definition is used:
\begin{align*}
&\mathrm{EE} = \\ &\frac{\textrm{Area spectral efficiency [bit/symbol/km}^2]}{\textrm{Transmit power + Circuit power per a.u.~[J/symbol/km}^2]}
\end{align*}
which yields to an EE metric measured in bit/J.\footnote{Some prior works have considered erroneous EE metrics in bit/J/Hz, by forgetting to scale both the numerator and the denominator with the symbol rate (or the bandwidth) that was explicitly (or implicitly) assumed in order to compute the noise power and the dimensionless SNR value. These works are not mentioned in this section for obvious reasons.}

Most of the early works that analyzed the EE, as defined above, have focused on the single-cell case wherein the interference from other cells is neglected. Examples in this context are \cite{Miao2013a,Hu2014a,Bjornson2013e,Ha2013a,Yang2013a,Mohammed2014a,Bjornson2014b,Bjornson2015a}. In particular, in \cite{Miao2013a} the authors consider the UL power allocation of multi-user MIMO systems and show that the EE is maximized when specific UEs are switched off. The UL was studied also in \cite{Hu2014a}, where the EE was shown to be a concave function of the number of BS antennas, $M$, and the UE rates. The DL has been investigated in \cite{Bjornson2013e,Ha2013a,Yang2013a}, whereof \cite{Bjornson2013e} and \cite{Ha2013a} show that the EE is a concave function of $M$, while a similar result is shown for the number of UEs, $K$, in \cite{Yang2013a}.
Unfortunately, in these works, the system parameters were optimized by means of simulations that (although useful) do not provide a complete picture of how the EE is affected by $M$ and $K$. Recently, \cite{Mohammed2014a} derived the optimal $M$ and $K$ for a given UL sum rate, while \cite{Bjornson2015a} derived the optimal $M$, $K$ and UE rates to maximize the EE jointly for the UL and DL. These two works utilize a realistic power consumption model where the dissipation in circuits depends non-linearly on the above parameters. Simple closed-form expressions for the EE-maximizing parameter values and their scaling behaviors are derived for zero-forcing (ZF) processing with perfect channel state information (CSI) and verified by simulations for other processing schemes under imperfect CSI.

The EE analysis of multi-cellular networks is much more involved than in the single-cell case due to the complicated network topology and the arising inter-cell interference. The simplest approach is to rely on heavy Monte Carlo simulations. Attempts in this direction can be found in \cite{Kurniawan2012} and \cite{Liu2013}. Unfortunately, Monte Carlo simulated results are often anecdotal since one cannot separate fundamental properties from behaviors induced by parameter selection.
Alternatively, simplified network topologies can be considered, such as the Wyner model \cite{Wyner1994a} or the symmetric grid-based deployment \cite{Bjornson2015a}. While attractive for its analytical simplicity, the Wyner model does not well capture the essential characteristics of real and practical networks \cite{Xu2011a}. Similarly, the symmetric grid-based models cannot capture the irregular structure of small-cell network deployments.

The need for developing tractable (yet reasonably accurate) models for future dense networks has increased the interest in random spatial models; more particularly, in using tools from stochastic geometry \cite{Haenggi-2009}, wherein the BS locations form a realization of a spatial point process---typically a Poisson point process (PPP). A major advantage of this approach is the ability of providing tractable expressions for key performance metrics such as the coverage probability and the average SE of the network. A few prior works have also derived EE-related performance metrics and showed how these depend on the BS and UE densities; for example, \cite{Cao2012a} compares the deployment of two types of single-antenna BSs, while \cite{Li2014a} studies the EE when multi-antenna BSs serve one UE each. In \cite{Soh2013}, the authors focus on the analysis and design of energy-efficient heterogeneous cellular networks (HetNets) through the deployment of small cells and sleeping strategies. The effect of cross-tier BS cooperation and clustering on the EE in the DL is studied in \cite{Nie2014}. In contrast to the UL analysis in our paper, the vast majority of work on the EE in multi-cell networks using stochastic geometry has focused on the DL. Furthermore, prior work usually aims at deriving closed-form SE and EE expressions, which often turn out to be involved and cumbersome to optimize, thereby providing little or no useful insights on the optimal system design.

\subsection{Major Contributions}

This paper considers the UL of a multi-cell multi-user MIMO network in which the BSs are distributed according to a homogenous PPP of intensity $\lambda$. Each BS is equipped with $M$ antennas and communicates with $K$ single-antenna UEs uniformly distributed within its coverage area. Coherent detection based on maximum ratio combining (MRC) is used at the BSs, based on CSI acquired from UL pilot signaling with each pilot symbol being reused in $1/\beta$ of the cells. In addition, we assume that the UEs are prone to practical transceiver hardware impairments and apply UL power control based on statistical channel inversion (with proportionality coefficient $\rho$), to resolve the near-far effects that otherwise would prevent communication \cite{Bjornson2016a}. Tools from stochastic geometry and classical statistics are used to compute a new lower bound of the average SE. This expression is then used to define the EE metric by also using the power consumption model developed in \cite{Bjornson2014b}, which not only accounts for the radiated power but also for the operating power consumption required by (among others) analog transceiver chains, digital processing, and backhaul infrastructure. 

Within the above conditions, an EE maximization problem is formulated under the assumption that a given average SE target per UE must be met with equality, to guarantee good service quality. This problem is solved analytically with respect to the tuple $\theta = (\beta,\lambda, \rho,M,K)$ of optimization variables. The resulting closed-form expressions provide valuable design insights on the interplay between the different system parameters and the various components of the power consumption model. We show analytically and numerically that the radiated power is negligible in most practical cases  and that massive MIMO configured BSs appear naturally since this technology can protect the desired signals from inter-cell interference and share the circuit power between many UEs.

\subsection{Outline and Notation}

The remainder of this paper is organized as follows. Section \ref{sec:system} introduces the system model for the network under investigation, including hardware impairments at the UEs and CSI estimation at the BSs. A new and tractable lower bound on the average SE is derived in Section \ref{sec:ASE}. It is used in Section \ref{sec:problem} to formulate the EE maximization problem and to obtain closed-form expressions for the EE-maximizing pilot reuse factor, BS density, and transmission power as well as the number of UEs and antennas per BS. The theoretical analysis is corroborated by numerical results in Section \ref{sec:simulations} while the major conclusions and implications of this paper are drawn in Section \ref{sec:conclusion}.

The following notation is used throughout the paper. The notation $\mathbb{E}\{ \cdot \}$ indicates the expectation with respect to a random variable, whereas $\| \cdot \|$ and $| \cdot |$ stand for the Euclidean norm and absolute value, respectively. We let $\vect{I}_{M}$ denote the $M \times M$ identity matrix, whereas we use $\mathcal{CN}(\cdot,\cdot)$ to denote a multi-variate circularly-symmetric complex Gaussian distribution. We use $ \mathbb{C}$, $ \mathbb{Z}_+$, and $\mathbb{R}$ to denote the sets of all complex-valued numbers, all positive integers, and all real-valued numbers, respectively. The Gamma function is denoted by $\Gamma(\cdot)$.

\section{System Model}

\label{sec:system}

\begin{figure}
\begin{center}
\includegraphics[width=0.95\columnwidth]{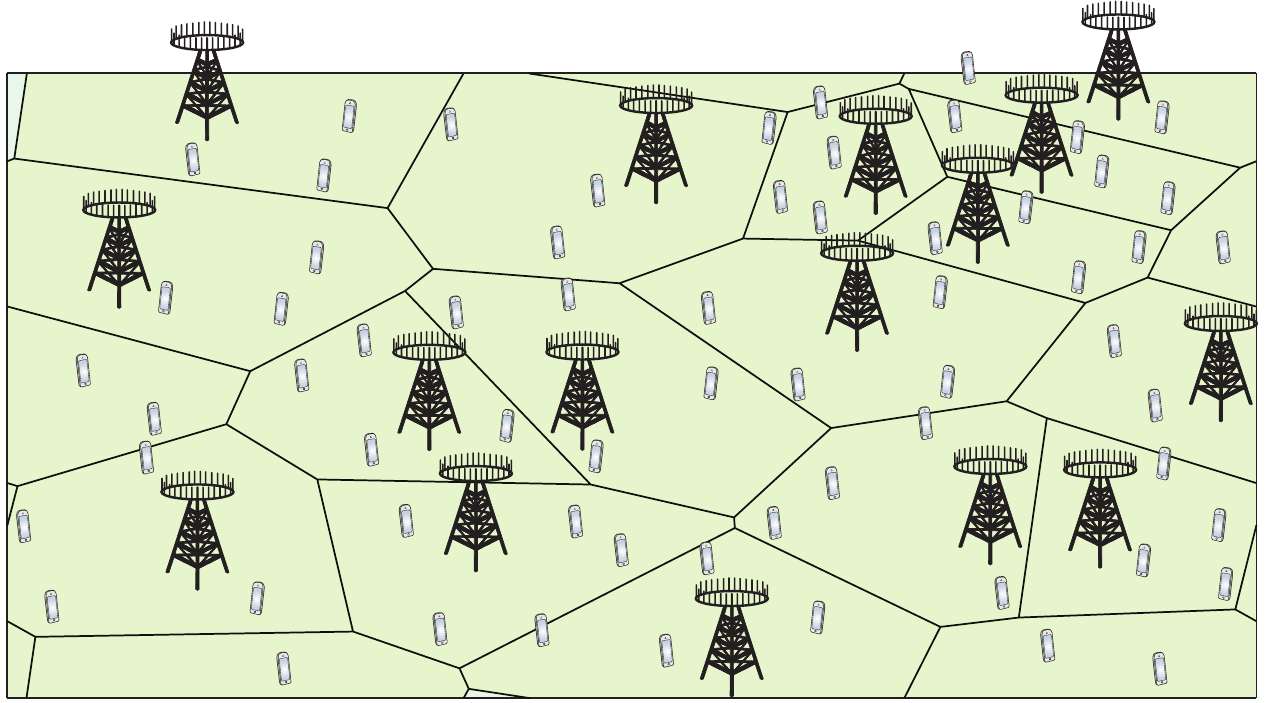}
\end{center}\vskip-3mm
\caption{Illustration of one realization of BS positions from the homogeneous PPP $\Phi_{\lambda}$ and of UEs uniformly distributed in the corresponding Poisson-Voronoi cells.} \label{figureVoronoi} \vskip-3mm
\end{figure}

We consider the UL of a cellular network that is designed to serve a dense heterogeneous distribution of UEs. This is modeled using the stochastic geometry framework adopted in \cite{Andrews2011a} in which the BSs are distributed spatially in $\mathbb{R}^2$ according to a homogeneous PPP $\Phi_{\lambda}$ of intensity $\lambda$ (measured in BSs per $\textrm{km}^2$). More precisely, this means that in any area of size $A$  (measured in $\textrm{km}^2$), the number of BSs is a Poisson distributed stochastic variable  with mean value $\lambda A$. The BSs are uniformly and independently distributed over the area. Each of them is equipped with an array of $M$ antennas and serves $K$ single-antenna UEs, which are selected at random from a (potentially) very large set of UEs in the cell since we treat $K$ as an optimization variable. We assume that each UE connects to its closest BS such that the coverage area of a BS is its Poisson-Voronoi cell; see Fig.~\ref{figureVoronoi} for an illustration. The UEs are assumed to be uniformly distributed in the Poisson-Voronoi cell of their serving BS. Note that the geographic locations of UEs and BSs are correlated under this model, such that small cells serve more UEs per $\textrm{km}^2$ than larger cells. We interpret this as a judicious network deployment where the BSs are matched to a heterogenous distribution of UEs.\footnote{Alternatively, the UEs could have been distributed according to an independent homogeneous PPP \cite{Li2014a}, leading to independent BS and UE locations. This is less sensible since smaller cells would on average have fewer UEs than larger cells, which contradicts the main principle of densifying networks mainly at places with high user loads.}

The translation invariance of PPPs allows us to perform statistical performance analysis for a \emph{typical UE} (located for example in the origin), which is statistically representative for any other UE in the network \cite{Baccelli2008a}. Assume that this typical UE has the arbitrary index $k$ and is connected to a BS that is called the \emph{typical BS} and is denoted as $\mathrm{BS}_0 \in \Phi_{\lambda}$. Then, the following basic properties hold (see for example \cite{Weber2010a}):

\begin{lemma} \label{lemma:distance-distribution}
The distance $d_{00k}$ from the typical UE to its serving BS is Rayleigh distributed as $d_{00k} \sim \mathrm{Rayleigh} \big( \frac{1}{\sqrt{2\pi \lambda}} \big)$. The BSs of the other cells form a homogenous PPP $\Psi_{\lambda} = \Phi_{\lambda} \setminus \{ \mathrm{BS}_0 \}$ in $\mathbb{R}^2 \setminus \{ \vect{x} \in \mathbb{R}^2 : \| \vect{x} \| < d_{00k} \}$.
\end{lemma}

Using the notation introduced in Lemma \ref{lemma:distance-distribution}, each interfering BS, $\mathrm{BS}_j \in \Psi_{\lambda}$, is situated at the geographical position determined by its spatial point $\mathrm{BS}_j \in \mathbb{R}^2$. Since in most cases we are interested only in the index of the interfering BS, the shorter notation $j \in \Psi_{\lambda}$ is used henceforth for $\mathrm{BS}_j \in \Psi_{\lambda}$.

\subsection{Channel Model and Power-Control Policy}

We consider a UL transmission protocol where the time-frequency resources are divided into blocks of $T_c$ seconds and $W_c$ Hz. This leaves room for a total number of $S = T_c W_c$ transmission symbols for pilot signaling and data. The channel response $\vect{h}_{ljk} \in \mathbb{C}^M$ between $\mathrm{BS}_l$ and UE $k$ in cell $j$ is modeled as block-fading such that $\vect{h}_{ljk}$ has a constant stochastic realization within a block and takes independent realizations across blocks. Since the number of BS antennas is treated as an optimization variable in this work, we need a channel model that is reasonable for both small and large arrays. For this purpose, we make use of a Rayleigh fading model as it well-matches non-line-of-sight measurements in both cases \cite{Gao2015a}. Hence, we consider Rayleigh fading channels with $\vect{h}_{ljk} \sim \mathcal{CN}(\vect{0}, \omega^{-1} d_{ljk}^{-\alpha} \vect{I}_M)$, where $d_{ljk}$ is the propagation distance (measured in km) and $\alpha> 2$ is the pathloss exponent. The parameter $\omega$ accounts for the pathloss at a reference distance of 1 km and can thus be used to model distance-independent propagation losses (such as wall penetration). 

Power control is an essential mechanism in the UL of any multi-user MIMO system; the finite dynamic range of analog-to-digital converters at the BSs otherwise creates near-far blockage where weak signals from distant UEs drown in stronger signals from nearby UEs. Statistical channel inversion is a convenient power-control policy \cite{Bjornson2016a}, where UE $i$ in cell $j$ uses the transmit power
\begin{equation}\label{p_{ji}}
p_{ji} = \rho  \omega d_{jji}^{\alpha}
\end{equation}
with $\rho \geq 0$ being a power-control coefficient to be designed.\footnote{We show later that the transmit power is low at EE-optimized operating points. To simplify the notation, we have therefore not specified any maximal transmit power constraints at the users.} The average transmit power at UE $i$ in cell $j$ will then be
\begin{equation} \label{p_{ji}_average}  
\mathbb{E}\{p_{ji}\} = \rho  \omega \mathbb{E}\{d_{jji}^{\alpha}\} = \rho  \omega \frac{ \Gamma({\alpha}/{2} +1 ) }{(\pi \lambda)^{\alpha/2}}
\end{equation}
by using the distance distribution from Lemma \ref{lemma:distance-distribution}.
A key feature of the statistical power-control policy is that it provides the same average effective channel gain of $\mathbb{E}\left\{{p_{ji} \|\vect{h}_{jji}\|^2} \right\} = M {\rho}$ to all UEs (irrespective of their locations) without requiring any rapid feedback mechanism to provide UEs with instantaneous CSI. Note that this does not mean that the signal-to-interference-plus-noise ratio (SINR) is the same for all UEs, since the interference varies in the network.

\subsection{Transceiver Hardware Impairments}

Any practical wireless transceiver is affected by hardware impairments; for example, unavoidable clock drifts in local oscillators, finite-precision digital-to-analog converters, amplifier non-linearities, non-ideal analog filters, etc. Such impairments can be partially mitigated by compensation algorithms \cite{wenk2010mimo}, but not fully removed. Although negligible at low signal-to-noise ratios (SNRs), hardware impairments characterize the maximal achievable SE at high SNRs \cite{Bjornson2013c}. The vast majority of prior works on the design of energy-efficient networks assumes ideal transceivers, although hardware impairments might have a fundamental impact also on the EE. Motivated by this fact, the distortion noise arising from hardware impairments is included in the system under investigation. We concentrate on UE impairments, which is expected to be the dominating effect since small BS arrays can use high-grade hardware and massive arrays are resilient to hardware impairments \cite{Bjornson2014a}.

Denote by $s_{ji} \in \mathbb{C}$ an arbitrary symbol transmitted by UE $i$ in cell $j$ and assume that it is normalized to unit power (i.e., $\mathbb{E}\{ | s_{ji}|^2 \}=1$). Similar to \cite{wenk2010mimo,Zhang2012a,Bjornson2013c}, the hardware impairments are modeled as a reduction of the desired signal power by a factor of $1-\epsilon^2$, with $0 \leq \epsilon < 1$, and by adding Gaussian distortion noise that carries the removed power. More precisely, $s_{ji} $ is replaced with $\sqrt{1-\epsilon^2}s_{ji} + e_{ji}$, where $e_{ji} \sim \mathcal{CN}(0,\epsilon^2)$ is the additive UL distortion noise caused at UE $i$ in cell $j$. In these circumstances,
the received signal $\vect{y}_0 \in \mathbb{C}^{M}$ at $\mathrm{BS}_0$ takes the form
\begin{equation} \label{eq:system-model-received}
\begin{split}
\vect{y}_0 &= \sum_{i=1}^{K} \sqrt{p_{0i}} (\sqrt{1-\epsilon^2}s_{0i} + e_{0i}) \vect{h}_{00i}  \\ &+ \sum_{j \in \Psi_{\lambda}}
\sum_{i=1}^{K} \sqrt{p_{ji}} (\sqrt{1-\epsilon^2}s_{ji} + e_{ji})  \vect{h}_{0ji}  + \vect{n}_{0}
\end{split}
\end{equation}
where $\vect{n}_{0} \sim \mathcal{CN}(0,\sigma^2 \vect{I}_M)$ models receiver noise with variance $\sigma^2$. The parameter $\epsilon$ is referred to as the \emph{level of impairments} and is tightly connected to the error vector magnitude (EVM), which is a common quality measure of transceivers.\footnote{The EVM is usually defined as the ratio between average distortion magnitude and signal magnitude, which becomes $\epsilon/\sqrt{1-\epsilon^2} \approx \epsilon$ with our notation.} The 3GPP LTE standards specify EVM requirements in the range $[0.05,0.175]$, where higher-order modulations are supported if the EVM is below the lower part of this range \cite{Holma2011a}. Note that the ideal hardware case is obtained by simply setting $\epsilon = 0$.

\subsection{Channel Acquisition}

As it is known, coherent processing of the received signal $\vect{y}_0$ requires knowledge of the UL channel vectors $\vect{h}_{00i}$ for $i=1,\ldots,K$. For this purpose, we assume that $B$ out of the $S$ symbols in each UL block are used for pilot transmission. This means that in the whole network there exists $B$ orthogonal pilot symbols that are shared among the cells. We assume that each BS picks $K \le B$ pilot symbols uniformly at random in each block, to avoid cumbersome pilot coordination. This means that on average $K/B$ of the cells reuse any given pilot symbol. We call $\beta = \frac{B}{K} \ge 1$ the pilot reuse factor such that $\beta K= B\le S$. In this setting, the pilot symbol, sent by the typical UE and received at the typical BS, is interfered with by the subset of cells in which there is another UE (still indexed by $k$ without loss of generality) reusing the same pilot symbol. This is modeled through a binary stochastic variable $\chi_{0kj} \in \{0,1\}$, where $\chi_{0kj}=1$ means that UE $k$ in cell $j$ uses the same pilot as the typical UE and thus causes pilot contamination. This occurs with probability $K/B$. Similarly, $\chi_{0kj}=0$ means that there is currently no pilot contamination from cell $j$, and it occurs with probability $1-K/B=1-1/\beta$.
Therefore, the typical UE $k$ transmits a single pilot symbol (e.g., $s_{0k}=1$) and $\mathrm{BS}_0$ receives the following signal:
\begin{equation} \label{eq:received-pilot}
\begin{split}
\vect{z}_{0k} &= \sqrt{p_{0k}} (\sqrt{1-\epsilon^2} + e_{0k}) \vect{h}_{00k} \\ &+ \sum_{j \in \Psi_{\lambda} } \chi_{0kj} \sqrt{p_{jk}} (\sqrt{1-\epsilon^2} + e_{jk}) \vect{h}_{0jk} + {\bf{n}}_{0}.
\end{split}
\end{equation}
We compute the minimum mean-squared error (MMSE) estimate of $\vect{h}_{00k}$ from the observation $\vect{z}_{0k}$.

 \begin{figure*}
 \begin{align} \tag{8}  \label{eq:effective-SINR}
\frac{ M(1-\epsilon^2)^2   }{  \!\!\!\!\Bigg( K + \fracSum{j \in \Psi_{\lambda}} \fracSumtwo{i=1}{K} 
\frac{d_{jji}^{\alpha}}{ d_{0ji}^{\alpha} } + \frac{\sigma^2}{  \rho}  \Bigg) \!\!\Bigg(  1   + \frac{1}{\beta} \fracSum{l \in \Psi_{\lambda} }   \frac{d_{llk}^{\alpha}}{ d_{0lk}^{\alpha} } + \frac{\sigma^2}{\rho} \Bigg) 
+ M (1-\epsilon^2) \Bigg( \frac{1}{\beta} \fracSum{j \in \Psi_{\lambda} }  \!\!
\left( \frac{ d_{jjk}^{\alpha} }{ d_{0jk}^{\alpha}} \right)^2  + \epsilon^2 \Bigg)  }
\end{align}
 \begin{align} \tag{10} \label{eq:average-SINR}
\frac{ M (1-\epsilon^2)^2  }{ 
\Big( K + \frac{\sigma^2}{  \rho}  \Big) \! \Big(  1   + \frac{ 2}{\beta( \alpha-2)}  + \frac{\sigma^2}{\rho} \Big) \! + \!
 \frac{ 2K}{\alpha-2} \! \left(  1  + \frac{\sigma^2}{\rho} \right) \! + \!
\frac{K}{\beta} \! \left( \frac{4}{(\alpha-2)^2 } +
\frac{1 }{ \alpha-1} \right) \!
+  M (1-\epsilon^2) \Big( \frac{ 1}{\beta( \alpha-1)} \!+\! \epsilon^2 \Big)  }
\end{align}
 \hrulefill
\end{figure*}

\begin{lemma}
The MMSE estimate of the typical UE's channel to its serving  $\mathrm{BS}_0$ is
\begin{equation} \label{eq:MMSE-estimator}
  \hat{\vect{h}}_{00k} = 
\frac{  \sqrt{ \frac{1-\epsilon^2}{\rho \omega d_{00k}^{\alpha}} }}{ 1   +  \sum_{j \in \Psi_{\lambda} }    \chi_{0kj}    \frac{d_{jjk}^{\alpha}}{ d_{0jk}^{\alpha} } + \frac{\sigma^2}{ \rho} } \vect{z}_{0k}.
\end{equation}
The estimation error $\Delta\vect{h}_{00k} = \vect{h}_{00k}  - \hat{\vect{h}}_{00k}$ is distributed as  $\Delta\vect{h}_{00k}  \sim\mathcal{CN}(\vect{0},\vect{C}_{00k})$, where the estimation error covariance matrix $\vect{C}_{00k} \in \mathbb{C}^{M \times M}$ is given by
\begin{equation} \label{eq:MMSE-error-cov}
\vect{C}_{00k} = \frac{1}{\omega d_{00k}^{\alpha}}  \left(1 - \frac{  1-\epsilon^2 }{ 1   +  \sum_{j \in \Psi_{\lambda} }    \chi_{0kj}   \frac{d_{jjk}^{\alpha}}{ d_{0jk}^{\alpha} } + \frac{\sigma^2}{\rho} }  \right) \vect{I}_M.
\end{equation}
\end{lemma}
\begin{IEEEproof}
This follows from standard results on MMSE estimation; see \cite[Chapter 15.8]{Kay1993a}.
\end{IEEEproof}

\section{Average Spectral Efficiency}
\label{sec:ASE}
We assume that MRC processing is used at the BSs for data recovery, since MRC is computationally efficient and performs well in both small cells (with $K=1$) and massive MIMO (with $M \gg K$). In particular, the symbol transmitted by the typical UE is detected at $\mathrm{BS}_0$ by correlating the received signal in
\eqref{eq:system-model-received} with the MMSE estimate $\hat{\vect{h}}_{00k}$; that is, $r_{0k} = \nu_{00k}\hat{\vect{h}}_{00k}^{\Htran}  \vect{y}_0$ is the received signal after MRC processing with $\nu_{00k} \in \mathbb{C}$ being a scaling factor (see Appendix A). Unfortunately, the ergodic capacity for a network (such as the one under investigation) in which only imperfect CSI is available at the receivers and the inter-cell interference is modeled as a shot-noise process is not known yet \cite{Weber2010a}. Hence, in this work the SE expression is obtained using classical achievable lower bounds on the ergodic capacity that are particularly common in the massive MIMO literature \cite{Marzetta2010a,Yang2013b,Bjornson2016a}. The following result holds for any given realization of $\Psi_{\lambda}$ and the UE locations.

\begin{lemma} \label{lemma:SINR-ideal}
With MRC processing and any given realization of $\Psi_{\lambda}$ and the UE locations, a lower bound on the ergodic capacity [bit/symbol/user] of the typical UE is
\begin{equation} \label{eq:ergodic-capacity}
\Big( 1 - \frac{\beta K}{S} \Big) \log_2 \! \left( 1 + \mathrm{SINR}_{0k}  \right)
 \end{equation}
 where the pre-log factor $\big( 1 - \frac{\beta K}{S} \big) $ accounts for pilot overhead and $\mathrm{SINR}_{0k}$, in \eqref{eq:effective-SINR} at the top of the page, is the effective SINR. \setcounter{equation}{8}
\end{lemma}
\begin{IEEEproof}
The proof is given in Appendix A and proceeds as follows. First, the lower-bounding approach from \cite{Jose2011b} and \cite{Medard2000a} is used to obtain the mutual information in the presence of imperfect CSI. Then, the statistical properties of the MMSE estimator and power-control policy are exploited to compute the average with respect to the channels in the presence of pilot contamination.
\end{IEEEproof}

The expression in Lemma \ref{lemma:SINR-ideal} holds for any $\beta\geq 1$ such that $\beta K \leq S$, since the pilot signals need to be contained in a coherence block. Observe also that $\beta K$ does not need to be an integer since an arbitrary $\beta K$ can be achieved by switching (for appropriate fractions of time) between the closest smaller integer, $\lfloor \beta K \rfloor$, and the closest larger integer, $\lceil \beta K \rceil$.

The average SE per UE can be obtained by taking the expectation of \eqref{eq:ergodic-capacity} with respect to the PPP $\Psi_{\lambda}$ and the UE locations. This would require heavy numerical evaluations of integrals. As a key contribution of this paper, in the following we provide a tractable and tight lower bound on the average SE.

\begin{proposition} \label{prop:average-SE}
If MRC is employed, a lower bound on the UL average SE [bit/symbol/user] is
\begin{equation} 
\underline {\mathrm{SE}} = \Big( 1 - \frac{\beta K}{S} \Big)\! \log_2 \! \left( 1 + \underline {\mathrm{SINR}} \right)\label{eq:average-SE}
\end{equation}
where $\underline {\mathrm{SINR}} $ is given in \eqref{eq:average-SINR} at the top of the page. \setcounter{equation}{10}
\end{proposition}
\begin{IEEEproof}
The proof is given in Appendix B and is based on Jensen's inequality, which leads to tractable expressions since only moments of the terms in \eqref{eq:effective-SINR} need to be computed.
\end{IEEEproof}

The tightness of the SE bound in Proposition \ref{prop:average-SE} is demonstrated later on by means of simulations (see Fig.~\ref{figureBSdensity}). The numerator of $\underline{\mathrm{SINR}} $ in \eqref{eq:average-SINR} scales with $M$ due to the array gain from coherent processing, but the hardware impairments cause the multiplicative loss $(1-\epsilon^2)^2$. The last term in the denominator scales with $M$ and also with the sum of $\frac{1}{\beta (\alpha-1)}$ and $\epsilon^{2}$. The former term is called coherent pilot contamination since it accounts for interference received from UEs that use the same pilot symbol as the typical UE, while the latter is due to the distortion noise emitted from the typical UE itself. Many of the interference terms in \eqref{eq:average-SINR} increase with $K$ since having more UEs lead to more transmit power per cell. Note that \eqref{eq:average-SINR} is independent of the BS density $\lambda$, due to the power-control policy in \eqref{p_{ji}}. The average transmit power per UE in \eqref{p_{ji}_average} is, however, proportional to $\lambda^{-\alpha/2}$ so that less power is needed (to sustain a fixed SNR per UE) as the network becomes denser.

\section{Problem Statement and Energy Efficiency Optimization}

\label{sec:problem}

As described in Section \ref{sec:intro}, we concentrate on the UL EE defined as the benefit-cost ratio between the area spectral efficiency (ASE) [bit/symbol/$\textrm{km}^2$] and the area power consumption (APC) [J/symbol/$\textrm{km}^2$]. Using the novel and tight lower bound from Proposition \ref{prop:average-SE}, in this work the ASE is obtained as
\begin{equation}
\mathrm{ASE} = \lambda K \, \underline{\mathrm{SE}}. \label{eq:def-ASE}
\end{equation}
To specify the APC, we begin by observing that with the adopted power-control policy the average radiated power per UE is
\begin{equation} 
\frac{S-(\beta K -1)}{S} \mathbb{E} \{ p_{ji} \} 
= \left( 1- \frac{\beta K - 1}{S} \right)  \rho  \omega \frac{ \Gamma({\alpha}/{2} +1 ) }{(\pi \lambda)^{\alpha/2}}
\end{equation}
where we have used \eqref{p_{ji}_average} and
the fact that each user transmits one pilot symbol and $S-\beta K$ data symbols per block.
Then, we observe that the APC must account not only for the radiated power, but also for the dissipation in analog hardware, digital signal processing, backhaul signaling, and other overhead costs (such as cooling and power supply losses).  A detailed and generic model that takes all these factors into account was recently proposed in  \cite{Bjornson2015a} and is such that (using the same notation as in \cite{Bjornson2015a} for simplicity) the APC is computed as
\begin{align} \notag
\mathrm{APC}  &=  \lambda \bigg( \left( 1- \frac{\beta K - 1}{S} \right) \frac{\rho \omega}{\eta}  \frac{ \Gamma({\alpha}/{2} +1 ) }{ (\pi \lambda)^{\alpha/2} } K  \\ & +\! \mathcal{C}_0 \! +\! \mathcal{C}_1 K \!+\! \mathcal{D}_0 M \!+\! \mathcal{D}_1 M K \bigg)  \!+\! \mathcal{A} \cdot \mathrm{ASE} \label{eq:def-APC}
\end{align}
where $\eta \in (0,1]$ is the linear power amplifier efficiency, $\mathcal{C}_0$ models the static power consumption at a BS, and $\mathcal{D}_0 M$ models the power consumption of the BS transceiver chains, which scales with the number of BS antennas.
Moreover, $\mathcal{C}_1 K + \mathcal{D}_1 M K$ models the power consumed at the UEs and also by the signal processing tasks at the BS, except for the coding and decoding that are proportional to the number of bits (with proportionality constant $\mathcal{A}$).  Notice that  \eqref{eq:def-APC} has an intuitive polynomial structure in $M$, $K$, and $\rho$. We refer to \cite{Bjornson2015a} for further modeling details.
The forthcoming analysis holds for any positive values of the above parameters, but some examples are later given in Table~\ref{table_parameters_hardware}.

The objective of this work is, for any given frame length ($S$), propagation parameters ($\alpha,\omega$), and hardware characteristics ($\eta,\epsilon, \mathcal{A},\mathcal{C}_0,\mathcal{C}_1,\mathcal{D}_0,\mathcal{D}_1$), to find the tuple of parameters $\theta = (\beta, \rho,\lambda,K,M)$ that solves the following constrained EE maximization problem:
\begin{equation} \label{eq:main-optimization-problem}
\begin{split}
\maximize{\theta \in \Theta} &\quad \mathrm{EE}(\theta) = \frac{\mathrm{ASE}(\theta)}{\mathrm{APC}(\theta)}\\
\mathrm{subject} \,\, \mathrm{to}  &   \quad \underline{\mathrm{SINR}} = \gamma 
\end{split}
\end{equation}
where $\Theta$ is the feasible parameter set defined as
\begin{equation}
\Theta = \{\theta: \, \rho\geq 0, \lambda \geq 0, \beta \geq 1, (M,K) \in \mathbb{Z}_+, K\beta \leq S\}
\end{equation}
with $K\beta \leq S$ being the upper limit on the pilot signaling overhead. The parameter $\gamma>0$ in \eqref{eq:main-optimization-problem} is used to impose an average SE constraint of $\log_2(1+\gamma)$ [bit/symbol/user], where the average is computed with respect to both BS and UE locations. This constraint is needed to obtain a network with an acceptable throughput, since unconstrained EE maximization often leads operating points with very low SEs per UE; this is later on illustrated in Fig.~\ref{figureBSdensity} where the EE increases as $\gamma$ decreases.

In the remainder of this section, we analyze the EE maximization problem in \eqref{eq:main-optimization-problem} to expose fundamental behaviors and to develop an algorithm for solving it.

\subsection{Feasibility}

\label{subsec:feasibility}

Due to the unavoidable inter-cell interference in cellular networks, the optimization problem \eqref{eq:main-optimization-problem} is only feasible for some values of $\gamma$. This feasible range is obtained as follows:
\begin{lemma}
The optimization problem \eqref{eq:main-optimization-problem} is feasible if \begin{equation} \label{eq:SINR-bound}
\gamma < \frac{S (\alpha -1) (1-\epsilon^2) }{ 1 + \epsilon^2 S (\alpha -1)}.
\end{equation}
\end{lemma}
\begin{IEEEproof}
Observe that $\underline {\mathrm{SINR}}$ in \eqref{eq:average-SINR} is a monotonically increasing function of $M$. Therefore, an upper limit can be computed by letting $M \rightarrow \infty$. This yields 
\begin{equation}
\underline {\mathrm{SINR}}
\xrightarrow{M \to \infty}
 \frac{(1-\epsilon^2)^2}{(1-\epsilon^2) \big( \frac{1}{\beta( \alpha-1)} + \epsilon^2 \big)},
\end{equation}
which is an increasing function of the optimization variable $\beta$. Under the constraint $\beta K \leq S$, the maximal value of $\beta$ is $S$ and this value is obtained for $K=1$. Putting all these facts together and using simple algebra lead to \eqref{eq:SINR-bound}.
\end{IEEEproof}

The above lemma shows that the maximal SINR level is limited only by the hardware impairments, through $\epsilon$, and by the severity of the pilot contamination, which is determined by the pathloss exponent $\alpha$ and the coherence block length $S$. These are the only limiting factors as $M \rightarrow \infty$.  Assume for example that $\epsilon = 0.05$ \cite{Bjornson2013c, Bjornson2014a} and consider the relatively conservative propagation parameters $\alpha=3$ and $S = 200$. For these numbers, the upper limit of the average SE per UE is $\log_2(1+199.5) \approx 7.65$, which is substantially higher than the SE of contemporary systems \cite{Holma2011a}. This means that the optimization problem  \eqref{eq:main-optimization-problem} is feasible in most cases of practical interest.

\begin{figure*}
\begin{align}  \label{eq:EEbeta}  \tag{23}
\mathrm{EE}(\beta^{\star}) = \frac{ K (1  \! -\! \frac{K}{S} \frac{B_1 \gamma}{M (1-\epsilon^2)^2  -B_2 \gamma} )  \log_2(1 \! + \! \gamma)}{ ( 1  \! + \!  \frac{1}{S} 
 \! -\!  \frac{K}{S} \frac{B_1 \gamma}{M (1-\epsilon^2)^2  -B_2 \gamma} ) \frac{K \rho \omega}{\eta}\frac{ \Gamma(\frac{\alpha}{2} +1 ) }{ (\pi \lambda)^{\alpha/2} }   \! + \!  \mathcal{C}_0  \! + \!  \mathcal{C}_1 K  \! + \!  \mathcal{D}_0 M  \! + \! \mathcal{D}_1 M K    \! + \!  \mathcal{A} K (1  \! -\! \frac{K}{S} \frac{B_1 \gamma}{M (1-\epsilon^2)^2  -B_2 \gamma} )  \log_2(1 \! + \! \gamma)  }
\end{align}
\begin{align} \label{eq:EE-limit}  \tag{25}
\mathrm{EE}_{\infty} =  \frac{ K (1 - \frac{K}{S} \frac{\bar{B}_1 \gamma}{M (1-\epsilon^2)^2  -\bar{B}_2 \gamma} )  \log_2(1+\gamma)}{  \mathcal{C}_0 + \mathcal{C}_1 K + \mathcal{D}_0 M + \mathcal{D}_1 M K   + \mathcal{A} K (1 - \frac{K}{S} \frac{\bar{B}_1 \gamma}{M (1-\epsilon^2)^2  -\bar{B}_2 \gamma} )  \log_2(1+\gamma)  }
\end{align}
\hrulefill
\end{figure*}

\subsection{Optimal Pilot Reuse Factor $\beta$}

We begin by deriving the optimal value of the pilot reuse factor $\beta$ when the other optimization variables are fixed.

\begin{theorem} \label{th:optimal-beta}
Consider any set of $\{\rho, \lambda, M, K\}$ for which the problem \eqref{eq:main-optimization-problem} is feasible. The SINR constraint in \eqref{eq:main-optimization-problem} is satisfied by selecting 
\begin{equation} \label{eq:beta-optimal}
\beta^\star = \frac{B_1 \gamma}{M(1-\epsilon^2)^2 - B_2 \gamma}
\end{equation}
where
\begin{align} \label{eq:B1}
B_1 &= \left( \frac{4K}{(\alpha-2)^2 } +
\frac{K +M (1-\epsilon^2) }{ \alpha-1} +  \frac{ 2 ( K + \frac{\sigma^2}{  \rho} )}{ \alpha-2} \right) \\
B_2 &=  \left( K + \frac{\sigma^2}{  \rho} +  \frac{ 2K}{\alpha-2} \right) \left(  1  + \frac{\sigma^2}{\rho} \right) + (1-\epsilon^2) \epsilon^2 M . \label{eq:B2}
\end{align}
\end{theorem}
\begin{IEEEproof}
By gathering the terms that contain $\beta$ and using \eqref{eq:average-SINR}, the SINR constraint can be rewritten as 
\begin{equation} \label{eq:gamma-equation}
\gamma = \frac{ M (1-\epsilon^2)^2  }{ B_1/\beta + B_2   }
\end{equation}
with $B_1$ and $B_2$ given by \eqref{eq:B1} and  \eqref{eq:B2}. We then obtain \eqref{eq:beta-optimal} by solving \eqref{eq:gamma-equation} for  $\beta$.
\end{IEEEproof}

The above theorem provides insights on how the EE-optimal pilot reuse factor $\beta^\star$ depends on the other system parameters. Firstly, recall that increasing $\beta$ translates into allocating a larger portion of each UL block for pilot transmission, so that each pilot symbol is on average only used in $1/\beta$ of the cells in the network. This leads to higher channel estimation accuracy and less coherent pilot contamination. Secondly, $\beta^\star$ is an increasing function of $B_1$ and also of $B_2$, since a larger $B_2$ makes the denominator smaller. Consequently, Theorem \ref{th:optimal-beta} shows that to guarantee a certain average SINR, $\beta^\star$ must increase with $K$. This is intuitive since more UEs per cell means more inter-cell interference, which can be partially suppressed by increasing the estimation accuracy and reducing the pilot contamination; namely, using a larger $\beta$. Similarly, $\beta^\star$ decreases with $\rho$ since higher transmit powers reduce the detrimental impact of noise, leading to higher estimation accuracy and to a more interference-limited regime. Moreover, $\beta^\star$ is a decreasing function of $M$ since an improved array gain makes the system less sensitive to interference and estimation errors. Increasing the pathloss exponent $\alpha$ leads to a smaller $\beta^\star$ (since $B_1$ and $B_2$ are reduced), which is natural since inter-cell interference decays more quickly.
Thirdly, the fact that $\beta \geq 1$ implies that we can only achieve values of $\gamma$ for which $ \frac{B_1 \gamma}{M(1-\epsilon^2)^2 - B_2 \gamma} \geq 1$, otherwise even $\beta=1$ would provide an SINR higher than $\gamma$.

\subsection{Optimal BS Density and Radiated Power}
Next, we optimize the BS density and radiated power. Plugging the optimal $\beta^\star$ from Theorem \ref{th:optimal-beta} into \eqref{eq:main-optimization-problem}, the EE maximization problem reduces to
\begin{align} \label{eq:main-optimization-problem-modified}
\maximize{\rho, \lambda \geq 0, \,\,  M,K \in \mathbb{Z}_+} &\quad \mathrm{EE}(\beta^{\star}) \\
\,\,\,\, \,\, \mathrm{subject} \,\, \mathrm{to} \,\,\, & \quad \frac{B_1 \gamma}{M(1-\epsilon^2)^2 - B_2 \gamma} \geq 1 \notag 
\\
& \quad  \frac{B_1 \gamma}{M (1-\epsilon^2)^2 -B_2 \gamma}  \leq \frac{S}{K} \notag
\end{align}
with $\mathrm{EE}(\beta^{\star}) $ given in \eqref{eq:EEbeta} at the top of the page. \setcounter{equation}{23}

The optimal values for the BS density $\lambda$ and the power-control coefficient $\rho$ are given in the following theorem.

\begin{theorem} \label{th:optimal-lambda}
Define $\rho = \lambda \tilde{\rho}$ for $\tilde{\rho}>0$ and consider any set of $\{\tilde{\rho},M,K\}$ for which the problem \eqref{eq:main-optimization-problem-modified} is feasible. Then, $\mathrm{EE}(\beta^{\star})$  is a monotonically increasing function of $\lambda$ and is maximized as $\lambda \to \infty$. The average transmit power per UE in \eqref{p_{ji}_average} will then go to zero.
\end{theorem}
\begin{IEEEproof}
The objective function $\mathrm{EE}(\beta^{\star})$ in \eqref{eq:EEbeta} is a monotonically increasing function of $\lambda$ since the transmit power term decreases with $\lambda$ proportionally to $\rho/\lambda^{\alpha/2} = \tilde{\rho}/\lambda^{\alpha/2-1}$.  The EE is also an increasing function of the factor $(1  \! -\! \frac{K}{S} \frac{B_1 \gamma}{M (1-\epsilon^2)^2  -B_2 \gamma} )$ and this expression also increases with $\lambda$ since $B_1$ and $B_2$ are decreasing with $\rho = \lambda \tilde{\rho}$. Therefore, $\mathrm{EE}(\beta^{\star})$ is maximized as $\lambda \to \infty$.
\end{IEEEproof}

This theorem proves that from an EE perspective it is preferable to have as high BS density as possible. This might be unexpected since smaller cells lead to more interfering UEs in the vicinity of each cell, but this issue is resolved by the assumed power-control policy that gradually reduces the transmit power as the BS density increases. The main consequence of letting $\lambda$ grow large is thus that the transmit power becomes negligible as compared to the circuit power in each cell. 

Clearly, an infinitely high BS density is infeasible in practice. However, the numerical results in Section \ref{sec:simulations} show that the asymptotic limit is almost achieved already at the modest density of $\lambda = 10$ BS/km$^2$. There are two factors that restrain the practical BS density: 1) the dimensionality of the BS equipment that limits the inter-BS distances; and 2) the UE density that limits the number of BSs that can serve $K$ UEs each. In ultra-dense networks, only a subset of all BSs are turned on at a given point in time; load balancing can be used to make sure that each active BS serves the most energy-efficient number of UEs, while the remaining ones are placed in sleep mode. This is further analyzed in Section \ref{subsec:given-UEdensity}.

\begin{figure*}
\begin{align} \label{eq:c1}  \tag{31}
\bar{c}^\star=  \frac{a_1  + a_3  + \sqrt{  a_1  a_3  + a_1^2+\frac{ a_1 a_2 a_4 }{a_5} + \frac{a_0  a_3 a_4 }{a_5} - \frac{a_0 a_1  a_4 }{a_5}  - \frac{a_0^2  a_3 a_4  }{a_2 a_5} + \frac{ a_0 a_1 a_3 }{a_2} + \frac{a_0 a_3^2}{a_2} } }{ a_2-a_0 }
\end{align} \hrulefill
\end{figure*}

\subsection{Optimal Number of Antennas and UEs per BS}
\label{subsec:opt-M-K}

By using Theorem \ref{th:optimal-lambda}, the EE maximization problem in \eqref{eq:main-optimization-problem-modified} further reduces to
\begin{align} \label{eq:main-optimization-problem-modified2}
\maximize{M,K \in \mathbb{Z}_+} &\quad  \mathrm{EE}_{\infty} \\
\,\,\,\, \,\, \mathrm{subject} \,\, \mathrm{to} \,\,\, & \quad  \frac{\bar{B}_1 \gamma}{M (1-\epsilon^2)^2 -\bar{B}_2 \gamma}  \geq 1 \notag \\
& \quad  \frac{\bar{B}_1 \gamma}{M (1-\epsilon^2)^2 -\bar{B}_2 \gamma}  \leq \frac{S}{K} \notag
\end{align}
where $\mathrm{EE}_{\infty}$ is given in \eqref{eq:EE-limit} at the top of the page and we have defined  \setcounter{equation}{25}
\begin{align} \label{eq:B1bar}
\bar{B}_1 &= K \left( \frac{4}{(\alpha-2)^2 } +
\frac{1}{ \alpha-1} +  \frac{ 2}{ \alpha-2} \right)  +
\frac{M (1-\epsilon^2) }{ \alpha-1} , \\
\bar{B}_2 &=  K\left(  1 +  \frac{ 2}{\alpha-2} \right) + (1-\epsilon^2) \epsilon^2 M . \label{eq:B2bar}
\end{align}
To find the optimal values for $M$ and $K$, an integer-relaxed version of \eqref{eq:main-optimization-problem-modified2} is first considered where $M$ and $K$ can be any positive scalars. The integer-valued solutions are then extracted from the relaxed problem. For analytic tractability, we replace $M$ with $\bar{c} = {M}/{K}$, which is the number of BS antennas per UE. For a given $\bar{c}$, the EE-maximizing value of $K$ is found as follows.

\begin{theorem} \label{th:optimal-K}
Consider the optimization problem \eqref{eq:main-optimization-problem-modified2} where $M$ and $K$ are relaxed to be real-valued variables. For any fixed $\bar{c} = {M}/{K}> 0$ such that the relaxed problem is feasible for some $K$, the EE is maximized by
\begin{align} \label{eq:optimal-K}
K^{\star} = \frac{ \sqrt{\left(G\mathcal C_{0}\right)^{2}+\mathcal{C}_{0} \mathcal D_{1}\bar c + \mathcal{C}_{0} G \left(\mathcal C_{1} + \mathcal D_{0}\bar c \right)}   - G\mathcal C_{0} }{\mathcal D_{1}\bar c+ G\left(\mathcal C_{1} + \mathcal D_{0}\bar c \right)}
\end{align} 
where
\begin{align} 
G = \frac{1}{S} \frac{ \left( \frac{4\gamma}{(\alpha-2)^2 } +
\frac{\gamma}{ \alpha-1} +  \frac{ 2\gamma}{ \alpha-2}  
 \right)  + \frac{ \gamma (1-\epsilon^2) }{ \alpha-1}  \bar{c} }{ 
 (1-\epsilon^2) \left( 1- (1+\gamma) \epsilon^2 \right) \bar{c}
  - \left(  1 +  \frac{ 2}{\alpha-2} \right) \gamma }.
\end{align}
\end{theorem}
\begin{IEEEproof}
If $\bar{c}$ is given, then we want to maximize (using the notation of the theorem and $\mathcal{R}=\log_2(1+\gamma)$)
\begin{align} \label{eq:utility-K}
 \frac{ K (1 - K G ) \mathcal{R}}{  \mathcal{C}_0 + ( \mathcal{C}_1+ \mathcal{D}_0 \bar c ) K  + \mathcal{D}_1 \bar c K^2   + \mathcal{A} K (1 - K G ) \mathcal{R}  } .
\end{align}
It is straightforward to show that \eqref{eq:utility-K} is a quasi-concave function of $K$ (e.g., using the approach in the proof of Lemma 3 in \cite{Bjornson2015a}). The maximizing value $K^{\star}$ is thus obtained by taking the first derivative of \eqref{eq:utility-K} and equating to zero. This leads to the expression in \eqref{eq:optimal-K}. The first constraint  in \eqref{eq:main-optimization-problem-modified2} is independent of $K$ and thus fulfilled if the problem was feasible. The solution also satisfies the second constraint, namely $K \leq {1}/{G}$, since the EE in \eqref{eq:utility-K} is negative for $K > {1}/{G}$ and thus is not maximized in thus range (and the zero objective function for $K=0$ would then be higher).
\end{IEEEproof}

Similarly, if $K$ is fixed, then the EE-maximizing value of  $\bar{c}$ is obtained as follows.

\begin{theorem} \label{th:optimal-c}
Consider the optimization problem \eqref{eq:main-optimization-problem-modified2} where $\bar{c} = {M}/{K}$ and $K$ are relaxed to be real-valued variables. For any fixed $K> 0$ such that the relaxed problem is feasible, the EE is maximized by $\bar{c}^\star$ in \eqref{eq:c1}, at the top of the page, if this solution satisfies the first inequality constraint in \eqref{eq:main-optimization-problem-modified2}. Otherwise, the EE is maximized by \setcounter{equation}{31}
\begin{align} \label{eq:c2} 
\bar{c}^\star= \frac{ \gamma \left( 1 +\frac{4}{(\alpha-2)^2 } + \frac{1}{ \alpha-1} +  \frac{ 4}{ \alpha-2}   \right)}{ (1-\epsilon^2) \left( 1- (1+\gamma) \epsilon^2 \right) - \frac{ \gamma (1-\epsilon^2) }{\alpha-1}  }.
\end{align}
The following parameters are used in \eqref{eq:c1} and \eqref{eq:c2}:
\begin{align}
a_0 &= \frac{ \gamma K (1-\epsilon^2) }{S(\alpha-1)} \\ 
a_1 &=  \frac{K}{S} \left( \frac{4\gamma}{(\alpha-2)^2 } + \frac{\gamma}{ \alpha-1} +  \frac{ 2\gamma}{ \alpha-2}   \right) \\
a_2 &=   (1-\epsilon^2) \left( 1- (1+\gamma) \epsilon^2 \right) \\
a_3 & = \left(  1 +  \frac{ 2}{\alpha-2} \right) \gamma \\
a_4 & =  \mathcal{C}_0 + \mathcal{C}_1 K \\
a_5 & = \mathcal{D}_0 K + \mathcal{D}_1 K^2.
\end{align}
\end{theorem}
\begin{IEEEproof}
Using the notation introduced in the theorem, the objective function in \eqref{eq:main-optimization-problem-modified2} reduces to
\begin{align} \label{eq:utility-c}
 \frac{ K (1 -  \frac{  a_0  \bar{c} + a_1  }{ a_2 \bar{c} - a_3} )  \log_2(1+\gamma)}{  a_4 + a_5 \bar{c}      + \mathcal{A} K (1 -  \frac{  a_0  \bar{c} + a_1  }{ a_2 \bar{c} - a_3} )   \log_2(1+\gamma)  } 
\end{align}
which can be easily shown to be a quasi-concave function of $\bar{c}$. The maximizing value $\bar{c}^{\star}$ is thus obtained by taking the first derivative of \eqref{eq:utility-c} and equating to zero. This yields the solution in \eqref{eq:c1}. Note that $\bar{c}^\star$ needs to satisfy the first constraint in \eqref{eq:main-optimization-problem-modified2}. If this is not the case with \eqref{eq:c1}, then the EE is monotonically increasing for all feasible $\bar{c}$ and the largest value is obtained when there is equality in the first constraint, which occurs at the value in \eqref{eq:c2}. Finally, we notice that $\bar{c}^\star$ satisfies the second constraint in \eqref{eq:main-optimization-problem-modified2} automatically because the maximum cannot give a negative EE.
\end{IEEEproof}

The above two theorems show how $K$ and $\bar{c}$ (and also $M=\bar{c}K$) are related at the EE-optimal points. 
It turns out that $K^{\star}$ decreases with $\bar{c}$, proportionally to $\sqrt{1/\bar{c}}$ when $\bar{c}$ grows large. Similarly,  $M^{\star} = \bar{c}^{\star}\!K$ is found to increase with $K$. The intuition is that more BS antennas should be deployed to control the interference when more UEs are served. In addition to this, the results of Theorems \ref{th:optimal-K} and \ref{th:optimal-c} reveal how the power consumption coefficients impact $K^{\star}$ and $M^{\star}$, respectively. In particular, from \eqref{eq:optimal-K} it is found that $K^{\star}$ increases with the static power consumption $\mathcal{C}_0$, while it decreases with $\mathcal{C}_1$, $\mathcal{D}_0$, and $\mathcal{D}_1$ that are the terms of the APC that increase with $K$ and $M$. Similarly, $M^{\star}$ increases with $\mathcal{C}_0$ and $\mathcal{C}_1$, but decreases with $\mathcal{D}_0$ and $\mathcal{D}_1$. The simple intuition behind these scaling behaviors is that more hardware should be turned on (i.e., BS antennas and UEs) only if the increase in circuit power has a marginal effect on the total APC. Similarly, with a larger static consumption we can afford to turn on more BS antennas and UEs since the relative power cost is lower.

Using Theorems \ref{th:optimal-K} and \ref{th:optimal-c}, we can devise an alternating optimization algorithm to solve the integer-relaxed EE maximization problem:

\begin{table*}[!t]
\renewcommand{\arraystretch}{1.3}
\caption{Simulation Parameters}
\label{table_parameters_hardware} \vskip-2mm
\centering
\begin{tabular}{|c|c|c||c|c|c|}
\hline
\bfseries System Parameter & \bfseries Symbol & \bfseries Value & \bfseries Hardware Parameter & \bfseries Symbol & \bfseries Value \\
\hline

Coherence block length & $S$ & $400$ & Coding/decoding/backhaul & $\mathcal{A}$ & $1.15 \cdot 10^{-9} \, \mathrm{[J/bit]}$ \\

Pathloss exponent & $\alpha$ & $3.76$ & Static power consumption & $\mathcal{C}_0$ & $10 \, \mathrm{W} \cdot \tau \,\, \mathrm{[J/symbol]}$ \\

Propagation loss at 1 $\textrm{km}$ & $\omega$ & $130 \, \mathrm{dB}$ & Circuit power per active user &  $\mathcal{C}_1$ & $0.1 \, \mathrm{W} \cdot \tau \,\, \mathrm{[J/symbol]}$ \\

Power amplifier efficiency & $\eta$ & $0.39$ & 
Circuit power per BS antenna & $\mathcal{D}_0$ & $0.2 \, \mathrm{W} \cdot \tau \,\, \mathrm{[J/symbol]}$ \\

Level of hardware impairments & $\epsilon$ & 0.05 & Signal processing coefficient & $\mathcal{D}_1$ & $1.56 \cdot 10^{-10} \, \mathrm{[J/symbol]}$ \\

Symbol time & $\tau$ & $\frac{1}{2 \cdot 10^7} \,\, \mathrm{[s/symbol]}$  & 
Noise variance & $\sigma^2$ & $10^{-20} \, \mathrm{[J/symbol]}$ \\

\hline
\end{tabular}
\end{table*}

\begin{enumerate}
\item Find a feasible starting point $(M,K)$ to \eqref{eq:main-optimization-problem-modified2};
\item Optimize $K$ for a fixed $M$ using Theorem \ref{th:optimal-K};
\item Optimize $M$ for a fixed $K$ using Theorem \ref{th:optimal-c};
\item Repeat 2)--3) until convergence is achieved.
\end{enumerate}

This algorithm converges since the EE has a finite upper bound and the EE increases monotonically in each iteration. In fact, it converges to the global optimum of the relaxed problem.

\begin{corollary}
The integer-relaxed version of the EE maximization problem \eqref{eq:main-optimization-problem-modified2} is quasi-concave. Hence, the alternating algorithm converges to the global optimal solution $(K^{\star \star},M^{\star \star}) \in \mathbb{R}^2$ to the integer-relaxed version of the problem.
\end{corollary}
\begin{IEEEproof}
The alternating optimization algorithm is guaranteed to converge to a local maximum according to \cite[Proposition 4]{Grippo2000a}. Suppose that there are multiple local maxima; for example, $(M_1,K_1)$ and $(M_2,K_2)$. By setting $M = \frac{M_2-M_1}{K_2-K_1} K + \frac{K_2 M_1- K_1 M_2}{K_2-K_1}$ and varying $K$, we can search on the line between these local maxima. For this choice of $M$, it is straightforward to show that the EE is a quasi-concave function of $K$. Hence, only one of the points can be a true local maximum, which is a contradiction to the existence of multiple local maxima; thus, the algorithm converges to the only local/global optimum of the problem.
\end{IEEEproof}

The real-valued solution $(K^{\star \star},M^{\star \star})$ obtained from the alternating optimization algorithm is a good starting point for finding the integer-valued global optimum to \eqref{eq:main-optimization-problem-modified2}. In particular, the quasi-concavity implies that the integer-solution is contained in a convex level set around $(K^{\star \star},M^{\star \star})$. In many cases, it is one of the four integer points obtained by respectively taking the floor and ceiling of $K^{\star \star}$ and $M^{\star \star}$. In general, the global optimum is obtained by searching through all integers in the vicinity of the real-valued solution, keeping in mind that the EE is quasi-concave in all directions.

To summarize, the original problem formulation in \eqref{eq:main-optimization-problem} has been solved through the following steps: $i$) by selecting $\beta$ to satisfy the SINR constraint; $ii$) by letting $\lambda \to \infty$ (which also makes $\rho \to 0$); $iii$) by devising an alternating optimization algorithm that provides the 
real-valued $M$ and $K$ that maximize the EE; $iv$) finally, by searching through the integer points in the vicinity of the real-valued solution, and capitalizing on the quasi-concavity. In the process of solving \eqref{eq:main-optimization-problem}, Theorems \ref{th:optimal-beta}--\ref{th:optimal-c} have also exposed the fundamental interplay among the optimization variables and how they depend on the hardware characteristics and propagation parameters.

\section{Numerical Examples}

\label{sec:simulations}

Numerical results are used in this section to validate the average SE expression provided in Proposition \ref{prop:average-SE} and the fundamental results established in Section \ref{sec:problem}. The results are obtained for the parameter setting summarized in Table \ref{table_parameters_hardware}, where we assume for instance a coherence block length of $S=400$ (e.g., $T_c=4$ ms and $W_c = 100$ kHz) and a bandwidth of 20 MHz (giving the symbol time $1/(2\cdot 10^7)$ s). The hardware parameters are inspired by a variety of prior works; for example, \cite{Bjornson2015a} and references therein. The simulations were performed using Matlab and the code is available for download at \url{https://github.com/emilbjornson/maximal-EE}, which enables reproducibility as well as simple testing of other parameter values.

\subsection{Optimizing the Energy Efficiency}

\begin{figure}
\begin{center}
\includegraphics[width=.95\columnwidth]{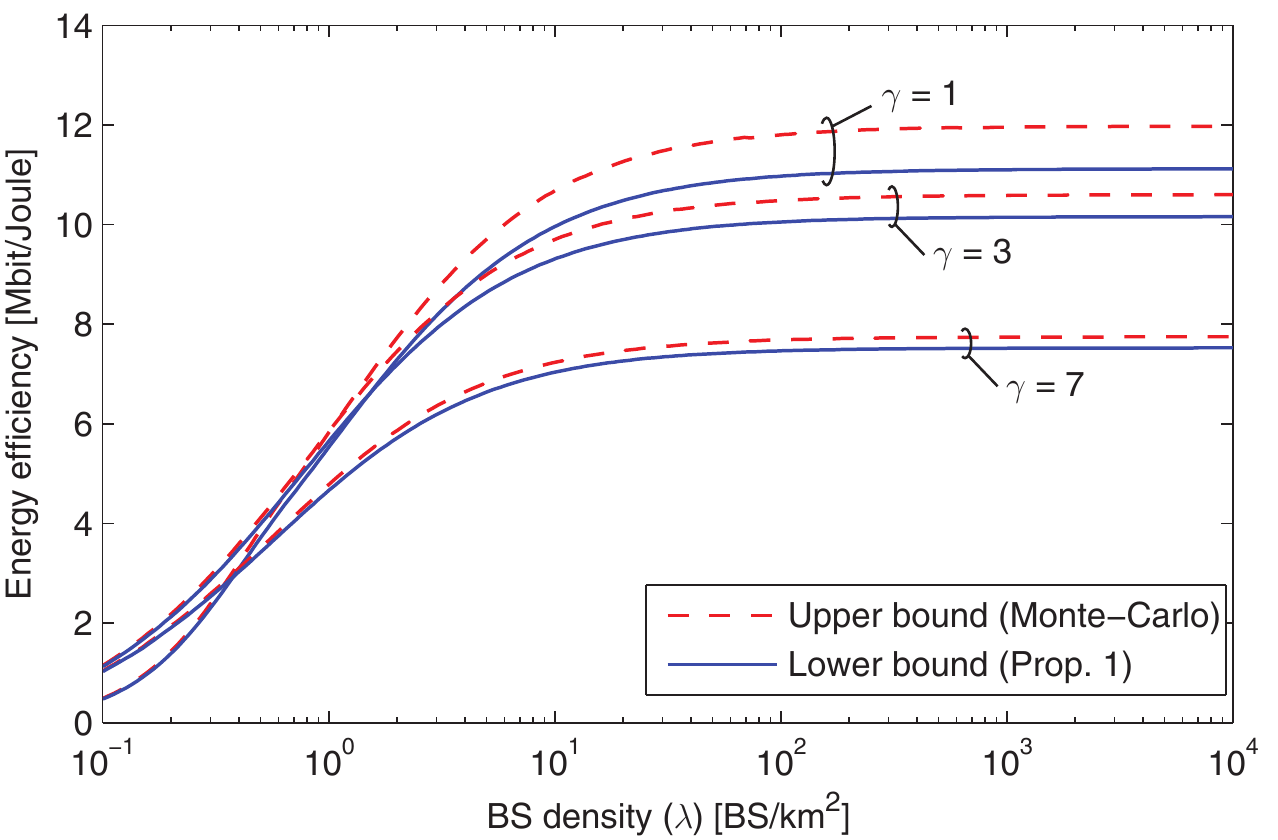}
\end{center}\vskip-3mm
\caption{Energy efficiency (in Mbit/J) as a function of the BS density in $\mathrm{BS}/\mathrm{km}^2$, for different SINR constraints. Results are obtained by both computing upper bounds from Monte-Carlo simulations and using the closed-form lower bound from Proposition\ref{prop:average-SE}.} \label{figureBSdensity} \vskip-3mm
\end{figure}

We begin by recalling that Theorem \ref{th:optimal-lambda} proves that the EE maximization problem in \eqref{eq:main-optimization-problem} is solved when the BS density is infinitely large; that is, $\lambda \to \infty$. We now illustrate how large $\lambda$ needs to be for applying this asymptotic result in practice. To this end, Fig.~\ref{figureBSdensity} shows the EE as a function of $\lambda$, and the other optimization variables are optimized numerically for each given value of $\lambda$. Three different SINR constraints are considered in Fig.~\ref{figureBSdensity}: $\gamma \in \{  1, \, 3, \, 7 \}$ which corresponds to the average SEs $\log_2(1+\gamma) \in \{  1, \, 2, \, 3 \}$. In all three cases, the EE is computed using both the lower bound on the average SE in Proposition \ref{prop:average-SE} and an upper bound obtained by averaging over the instantaneous SE derived in Lemma \ref{lemma:SINR-ideal} using Monte Carlo simulations.\footnote{This is an upper bound since we only consider an average of 1000 closest interfering BSs, while the exact result requires an infinite number of interferers in $\mathbb{R}^2$.}

Several important observations can be made from the results presented in Fig.~\ref{figureBSdensity}. Firstly, there is only a small gap (that reduces as $\gamma$ takes larger values) between the lower and upper bounds, and the curves behave exactly the same for any value of $\gamma$. This validates the accuracy of the SE expression provided in Proposition \ref{prop:average-SE}. Secondly, the EE can be greatly improved by increasing the BS density, meaning that small cells are a promising solution for maximal EE deployment. However, the gain from increasing the BS density saturates in the interval from $\lambda = 10$ to $\lambda = 100$  BS/km$^2$, which roughly corresponds to an average inter-BS distance of 100--315 meters. This is not small compared to contemporary urban deployments. In other words, EE maximization based on letting $\lambda \rightarrow \infty$ is expected to give representative results in most practical dense deployments. Thirdly, we note that the EE decreases as $\gamma$ increases, which is why it is important to specify a target SINR; otherwise the EE maximizing operating point might be very spectrally inefficient and therefore useless from a practical user service perspective.

\begin{figure}
\begin{center}
\includegraphics[width=.95\columnwidth]{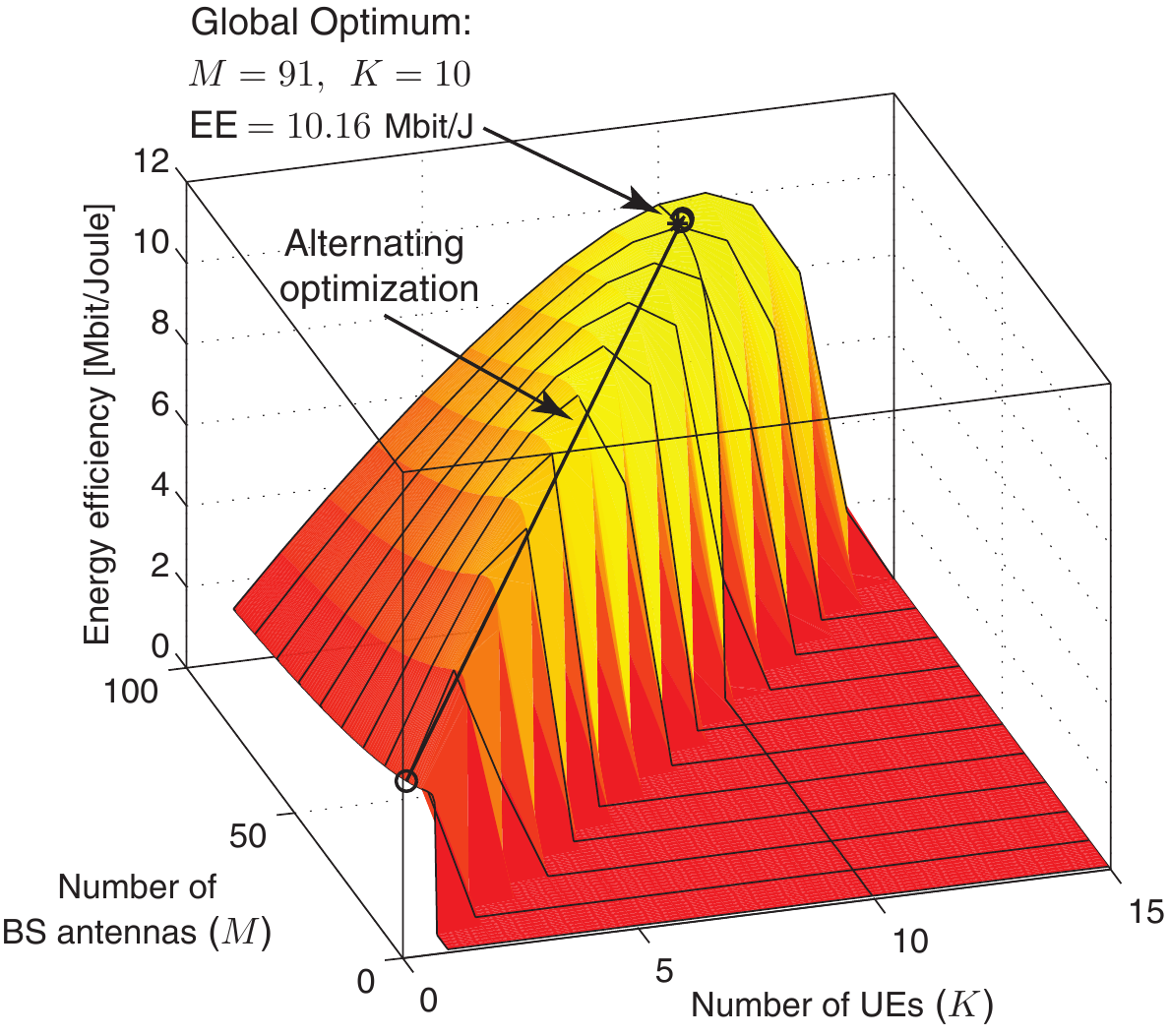}
\end{center}\vskip-3mm
\caption{Energy efficiency (in Mbit/J) for $\gamma = 3$. The global optimum is star-marked, while the convergence
of the alternating algorithm from Section \ref{subsec:opt-M-K} is indicated with circles.} \label{figure3d}  \vspace{-0.2cm}
\end{figure}

We proceed by further studying the impact of varying $M$ and $K$. To this end, we let $\lambda \rightarrow \infty$ as prescribed by Theorem \ref{th:optimal-lambda} and fix the SINR constraint at $\gamma = 3$ or, equivalently, the average SE per data symbol at $\log_2(1+\gamma) = 2$. Fig.~\ref{figure3d} shows the EE lower bound as a function of $M$ and $K$ when $\beta$ is optimized according to \eqref{eq:beta-optimal}. The global EE maximum gives an EE of $10.156$ Mbit/J and is achieved by $(M^\star,K^\star) = (91,10)$ using the pilot reuse factor $\beta^\star = 7.08$. Interestingly, this is a configuration that falls within the class of massive MIMO setups \cite{Marzetta2010a,Larsson2014a,Vieira2014a}. The intuition behind this result is that the strong inter-cell interference in dense deployments is efficiently mitigated by MRC when the BSs are equipped with many antennas and also by using a substantial pilot reuse factor (to protect against channel estimation errors and pilot contamination).

The results obtained with the alternating optimization algorithm from Section \ref{subsec:opt-M-K} are also shown in Fig.~\ref{figure3d}. The initialization point was set to $(M,K) = (20,1)$. As seen, the algorithm converges after three iterations to the real-valued solution $(M^{\star \star},K^{\star \star}) = (91.6,10.1)$ with an EE of $10.157$ Mbit/J. This real-valued operating point gives only a 0.009\% higher EE than the  integer-valued solution, thus showing that the EE performance is quite flat around the global optimum.

\begin{figure}
\begin{center}
\includegraphics[width=.7\columnwidth]{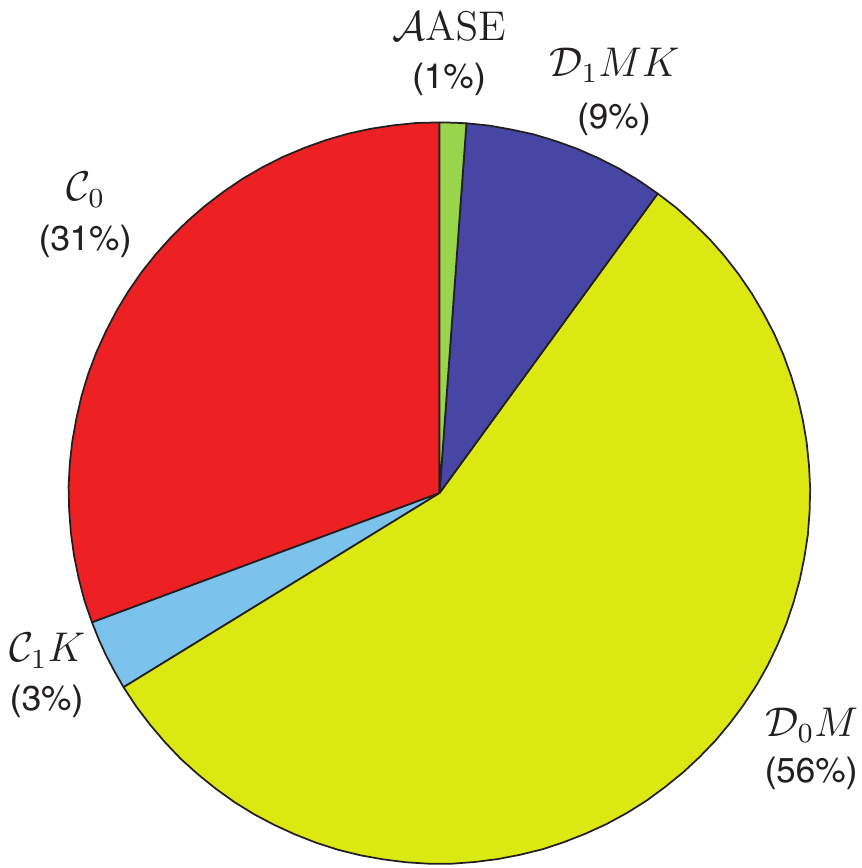}
\end{center}\vskip-3mm
\caption{Relative size of each term in the APC of \eqref{eq:def-APC} at the global optimum from Fig.~\ref{figure3d}.} \label{figurePieDiagram} \vspace{-0.2cm}
\end{figure}

To study the global optimum even further, the pie diagram in Fig.~\ref{figurePieDiagram} shows the relative size of each term in the APC of \eqref{eq:def-APC} (the transmit power is not shown since it takes negligible values). The dominating terms are the static power consumption $\mathcal{C}_0$ and the consumption $\mathcal{D}_0 M$ of the BS transceiver chains. These seem to be the main factors to improve in order to make the hardware more energy-efficient. Observe that the above results might be different for other values of $\mathcal{C}_0$, $\mathcal{C}_1$, $\mathcal{D}_0$, and $\mathcal{D}_1$.

\subsection{Impact of Transceiver Hardware Impairments}

\begin{figure}
\begin{center}
\includegraphics[width=.95\columnwidth]{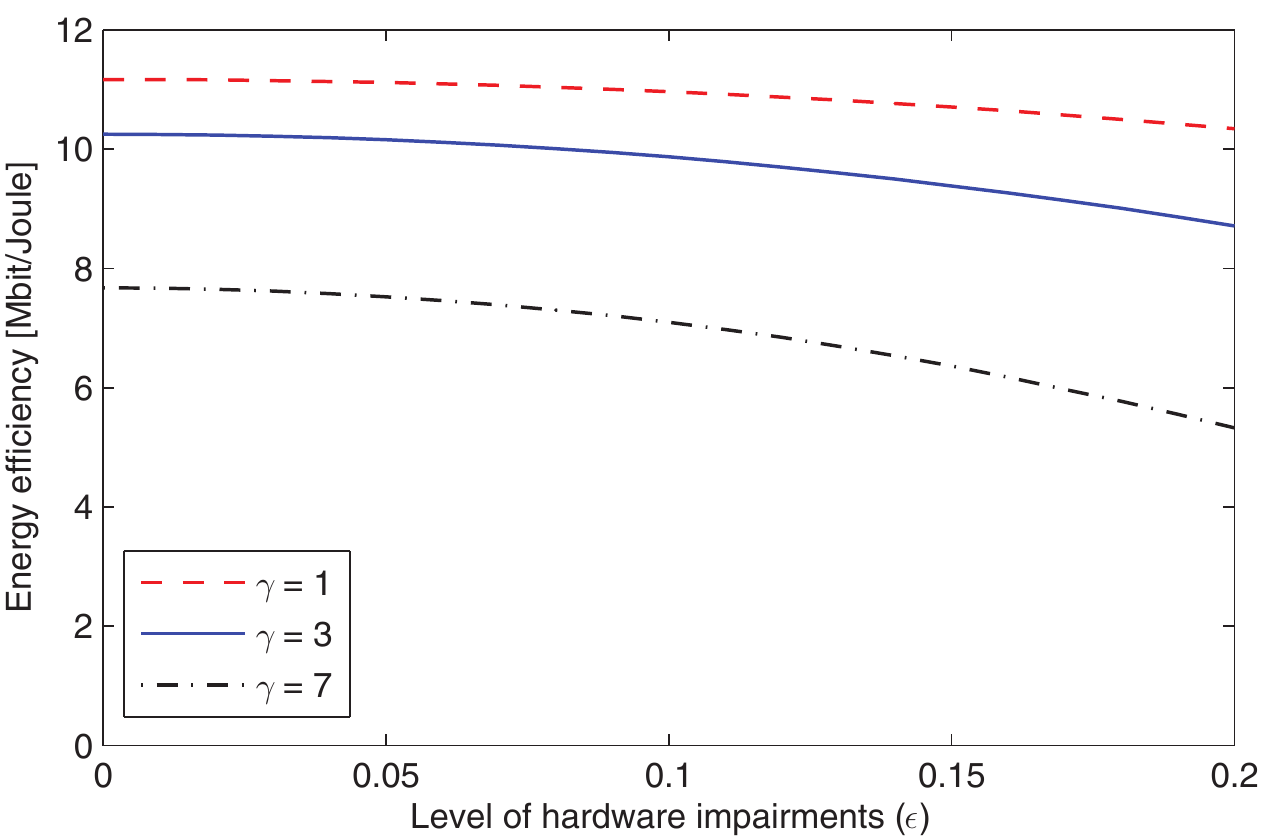}
\end{center}\vskip-3mm
\caption{Energy efficiency (in Mbit/J) as a function of the level of transceiver hardware impairments, $\epsilon$, for different SINR constraints.} \label{figureImpairments} \vspace{-0.2cm}
\end{figure}

Next, we consider the saturation regime where $\lambda \to \infty$ and exemplify the impact of transceiver hardware impairments on the EE. Fig.~\ref{figureImpairments} shows the EE as a function of $\epsilon$, the level of hardware impairments. As expected, the EE decreases with $\epsilon$ since the desired signal power in \eqref{eq:average-SINR} decays as $(1-\epsilon^2)^2$. The loss is marginal for $\gamma = 1$, but it can be relatively large when $\gamma$ increases. This is in line with the previous results in \cite{Bjornson2013c}, wherein it is shown that hardware impairments greatly affect the channel capacity in the high SNR regime (i.e., for large $\gamma$) while their impact is negligible in the low SNR regime. The results of Fig.~\ref{figureImpairments} indicate that for the investigated network the EE loss due to hardware impairments is negligible for $\epsilon \leq 0.1$ when $\gamma \in \{1, \, 3 \}$. Since these values correspond to the operating points that give the highest EE (see Fig.~\ref{figureBSdensity}), we may conclude that modest levels of hardware impairments have a negligible impact on networks designed for high EE.

\subsection{EE Maximization for a Given UE Density}
\label{subsec:given-UEdensity}

So far, the BSs have been deployed to match an unlimited heterogeneous UE distribution. In particular, we have assumed that each BS serves $K$ UEs such that a BS density of $\lambda$ BS/km$^2$ corresponds to an average of $K \lambda$ UEs per km$^2$. Hence, as the BS density grows large we also let the average UE density grow large. To validate the plausibility of this model, suppose that we instead deploy the BSs to match a fixed average UE density of $\mu$ UE/km$^2$. Mathematically, this amounts to solving \eqref{eq:main-optimization-problem} with the additional constraint
\begin{equation} \label{eq:UE-density-constraint}
\mu = K \lambda. 
\end{equation}
We will study how such an extra constraint affects the results, taking into account that future average UE densities from $\mu=10^2$ UE/km$^2$ (in rural areas) to $\mu=10^5$ UE/km$^2$ (in shopping malls) have been predicted in the METIS project \cite{METIS_D11_short}.

\begin{figure}[t!]
\begin{center}
\includegraphics[width=.95\columnwidth]{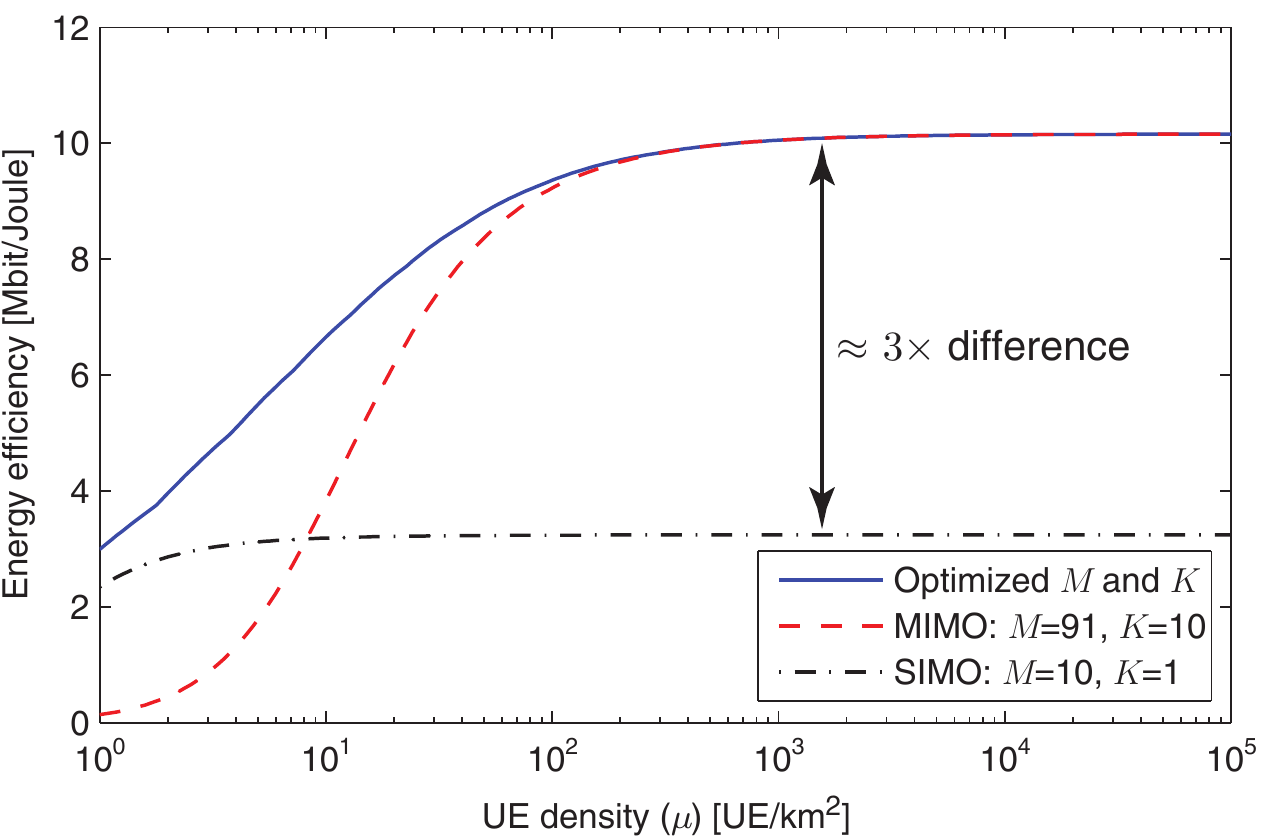}
\end{center}\vskip-3mm
\caption{Energy efficiency (in Mbit/J) vs. the UE density $\mu$. The EE is optimized according to \eqref{eq:main-optimization-problem} with the extra constraint $\mu = K \lambda$, or only with respect to $(\lambda,\beta,\rho)$ for given $M$ and $K$.}
\label{figureUEdensity_EE} \vspace{-0.2cm}
\end{figure}

\begin{figure}[t!]
\begin{center}
\includegraphics[width=.95\columnwidth]{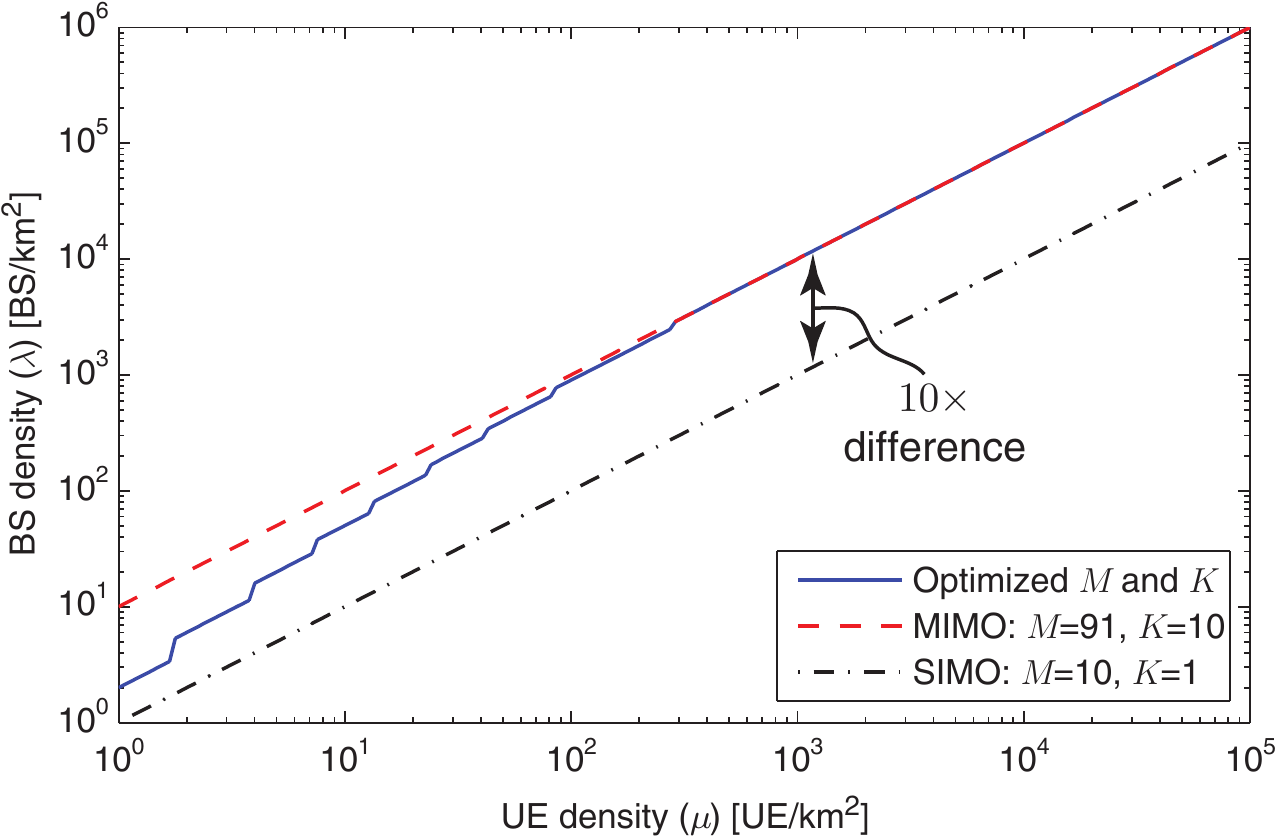}
\end{center}\vskip-3mm
\caption{Optimized BS density (in $\mathrm{BS/km}^2$) vs. the UE density (in $\mathrm{UE/km}^2$). The system is optimized in the same way as in Fig.~\ref{figureUEdensity_EE}.} \label{figureUEdensity_BS}
 \vspace{-0.2cm}
\end{figure}

Fig.~\ref{figureUEdensity_EE} shows the EE has a function of the UE density $\mu $ for the average SINR level $\gamma = 3$, while Fig.~\ref{figureUEdensity_BS} shows the corresponding BS density. The design parameters $M$, $K$, $\beta$, $\lambda$, and $\rho$ are optimized numerically according to \eqref{eq:main-optimization-problem}, with the additional constraint \eqref{eq:UE-density-constraint}. Two reference cases are also considered: single-user single-input multiple-output (SIMO) transmission with $(M,K) = (10,1)$; and massive multi-user MIMO transmission with $(M,K) = (91,10)$, which was shown in Fig.~\ref{figure3d} to be  optimal as $\lambda \to \infty$. Clearly, only $\beta$, $\lambda$, and $\rho$ are optimized in the two reference cases.

Several important observations can be made from the results presented in Figs.~\ref{figureUEdensity_EE} and~\ref{figureUEdensity_BS}. Firstly, the EE level becomes independent of the UE density when $\mu$ is large enough; the saturation occurs for $\mu \geq 100$ in the optimal and the fixed massive MIMO cases, while it occurs for $\mu \geq 2$ in the single-user SIMO case. The future UE density range $\mu \in [10^2, 10^5]$ predicted in \cite{METIS_D11_short} is fully covered in these saturation regimes. In all practical relevant cases we can, according to Fig.~\ref{figureUEdensity_BS}, simply scale the BS density linearly with the UE density, while using the same number of BS antennas and UEs per BS. Similarly, one can turn on and off BSs to account for variations in the UE load. Secondly, the fixed massive MIMO configuration $(M,K) = (91,10)$ achieves the maximal EE in the saturation regime, while it is inefficient as compared to single-user MISO for low UE densities (e.g., for $\mu \leq 8$). In contrast, single-user SIMO transmission performs reasonably well at low UE densities, but saturates earlier and at an EE level that is $3.14 \times$ lower than the maximal EE. More importantly, Fig.~\ref{figureUEdensity_BS} shows that the single-user case requires a $10 \times$ higher BS density in the saturation regime, which might greatly reduce the deployment cost.

In summary, the transmission power appears to be negligible, as compared to the circuit power, for networks that can handle future UE densities. By adding massive multi-user MIMO capability to the BSs, the EE can be increased by a few hundred percentages, and the even more substantial benefit is that it can reduce the BS density with an order of magnitude; this is likely to be a key property to achieve a cost and energy efficient network densification.

\section{Conclusion}
\label{sec:conclusion}

Network densification is the key for achieving high EE in cellular networks, and can be realized by either having many BSs or many antennas per BS. In order to find the optimal densified network configuration, we formulated an UL EE maximization problem under the assumption of a stochastic BS deployment based on Poisson point processes. By deriving a new lower bound on the average SE in the network and using a state-of-the-art power consumption model, the EE expression became tractable and was maximized analytically with respect to the density of BSs, the transmit power levels, the number of BS antennas and users per cell, and the pilot reuse factor. The closed-form expressions provided general guidelines on the optimal operating regimes and exposed the fundamental interplay among the optimization variables, hardware characteristics, and propagation environment. The analysis focused on the UL, but similar results are expected in the DL due to the UL/DL duality concept.

The analysis shows that reducing the cell size is undoubtable the way towards high EE, but the positive effect of increasing the BS density saturates when the circuit power dominates over the transmission power. A further leap in EE can typically be achieved by adding extra BS antennas to multiplex several UEs per cell; the numerical examples resulted in an EE maximum with 91 BS antennas and 10 UEs, which is a massive MIMO setup. The EE gains come from suppressing intra-cell interference and by sharing the circuit power costs among multiple UEs. We stress that a hundred antennas can be deployed also at a small-size BS, since we consider simple dipoles and possibly higher frequencies than in contemporary networks. Moreover, the analysis shows that a large pilot reuse factor can protect against inter-cell interference and can be tailored to guarantee a certain average SE.

While this paper focused on the ratio between spectral efficiency and energy costs, it is straightforward to also include the deployment cost, site renting, and other economical factors; the annual depreciation expense can be turned into an energy-equivalent form (by multiplying with the energy cost in J/Euro) and incorporated in the coefficients $\mathcal{C}_0$, $\mathcal{C}_1$, and $\mathcal{D}_0$ in \eqref{eq:def-APC}. This is likely to mainly increase the constant term $\mathcal{C}_0$, thus pushing the optimal operating point towards having fewer BSs, with more antennas and UEs per cell.

\begin{figure*}
 \begin{align} \notag
&\mathbb{E} \{ \mathrm{SINR}_{0k}^{-1} \} \\ \notag
& = \frac{ \mathbb{E} \left\{ \left( K + \fracSum{j \in \Psi_{\lambda}} \fracSumtwo{i=1}{K} 
\frac{d_{jji}^{\alpha}}{ d_{0ji}^{\alpha} } + \frac{\sigma^2}{  \rho}  \right) \left(  1   + \frac{1}{\beta} \fracSum{l \in \Psi_{\lambda} }   \frac{d_{llk}^{\alpha}}{ d_{0lk}^{\alpha} } + \frac{\sigma^2}{\rho} \right) \right\}
+ M (1-\epsilon^2) \left(  \mathbb{E} \left\{ \frac{1}{\beta} \fracSum{j \in \Psi_{\lambda} }  
\left( \frac{ d_{jjk}^{\alpha} }{ d_{0jk}^{\alpha}} \right)^2 \right\} + \epsilon^2 \right) }{ M (1-\epsilon^2)^2  } \\ \notag
&=   \frac{1}{ M (1-\epsilon^2)^2} \left(
\bigg( K + \frac{\sigma^2}{  \rho}  \right) \left(  1  + \frac{\sigma^2}{\rho} \right) 
+\mathbb{E} \left\{ \fracSum{j \in \Psi_{\lambda}} \fracSumtwo{i=1}{K} 
\frac{d_{jji}^{\alpha}}{ d_{0ji}^{\alpha} } \right\} \left(  1  + \frac{\sigma^2}{\rho} \right) +   \left( K  + \frac{\sigma^2}{  \rho}  \right) \mathbb{E} \bigg\{ \frac{1}{\beta} \fracSum{l \in \Psi_{\lambda} }   \frac{d_{llk}^{\alpha}}{ d_{0lk}^{\alpha} }  \bigg\} \\
&+ 
 \mathbb{E} \bigg\{ \frac{1}{\beta}  \bigg(  \fracSum{j \in \Psi_{\lambda}} \fracSumtwo{i=1}{K} 
\frac{d_{jji}^{\alpha}}{ d_{0ji}^{\alpha} }  \bigg) \bigg(   \fracSum{l \in \Psi_{\lambda} }   \frac{d_{llk}^{\alpha}}{ d_{0lk}^{\alpha} }  \bigg) \bigg\}
+ M (1-\epsilon^2) \mathbb{E} \bigg\{ \frac{1}{\beta} \fracSum{j \in \Psi_{\lambda} }  
\left( \frac{ d_{jjk}^{\alpha} }{ d_{0jk}^{\alpha}} \right)^2 \bigg\} + M (1-\epsilon^2) \epsilon^2 \bigg).  \label{eq:inverse-SINR} \tag{51}
\end{align} \hrulefill
\end{figure*}

\appendices

\section*{Appendix A: Proof of Lemma \ref {lemma:SINR-ideal}}
\label{app:lemma-SINRs}

Let $\vect{v}_{00k} = \nu_{00k} \hat{\vect{h}}_{00k}$ denote the MRC detector. Then, the received signal is given by
\begin{align} \notag
& \vect{v}_{00k}^{\Htran} \vect{y}_0 = \sum_{i=1}^{K} \sqrt{p_{0i}} (\sqrt{1-\epsilon^2}s_{0i} + e_{0i}) \vect{v}_{00k}^{\Htran} \vect{h}_{00i}  \\ & + \sum_{j \in \Psi_{\lambda}} \notag
\sum_{i=1}^{K} \sqrt{p_{ji}} (\sqrt{1-\epsilon^2}s_{ji} + e_{ji})  \vect{v}_{00k}^{\Htran} \vect{h}_{0ji}  + \vect{v}_{00k}^{\Htran} \vect{n}_{0} \\ \notag &= \sqrt{p_{0k}} \sqrt{1-\epsilon^2}s_{0k} \mathbb{E} \{ \vect{v}_{00k}^{\Htran} \vect{h}_{00k} \} + 
\sqrt{p_{0k}} e_{0k} \vect{v}_{00k}^{\Htran} \vect{h}_{00k} \\ \notag &+ \sqrt{p_{0k}} \sqrt{1-\epsilon^2}s_{0k} \left( \vect{v}_{00k}^{\Htran} \vect{h}_{00k} - \mathbb{E} \{ \vect{v}_{00k}^{\Htran} \vect{h}_{00k} \} \right)   &  \\ \notag &+ 
 \sum_{i \neq k} \sqrt{p_{0i}} (\sqrt{1-\epsilon^2}s_{0i} + e_{0i}) \vect{v}_{00k}^{\Htran} \vect{h}_{00i}  + \notag \\ &+ \sum_{j \in \Psi_{\lambda}}
\sum_{i=1}^{K} \sqrt{p_{ji}} (\sqrt{1-\epsilon^2}s_{ji} + e_{ji})  \vect{v}_{00k}^{\Htran} \vect{h}_{0ji}  + \vect{v}_{00k}^{\Htran} \vect{n}_{0}. \label{eq:processed-signal}
\end{align}
The ergodic capacity of this channel is defined as the maximal mutual information, which unfortunately cannot be computed exactly under imperfect CSI. However, we can compute a rigorous lower bound using the well-established bounding technique from \cite{Jose2011b} and \cite{Medard2000a}. More precisely, the first term in \eqref{eq:processed-signal} is treated as the only desired signal while the other terms are uncorrelated with the first one and treated as worst-case Gaussian noise in the detection (a worst-case assumption). This leads to the mutual information $\log_2(1+\mathrm{SINR}_{0k})$ where $\mathrm{SINR}_{0k}$ is the ratio between the desired power and the ``noise'' power:
\begin{align} \notag
&\mathrm{SINR}_{0k} = p_{0k} (1-\epsilon^2) |\mathbb{E} \{ \vect{v}_{00k}^{\Htran} \vect{h}_{00k} \}|^2 \\ & \notag \Big/ \Big(
p_{0k} (1-\epsilon^2) \left( \mathbb{E} \{ | \vect{v}_{00k}^{\Htran} \vect{h}_{00k} |^2 \} - |\mathbb{E} \{ \vect{v}_{00k}^{\Htran} \vect{h}_{00k} \}|^2 \right)   \\ &+
p_{0k} \epsilon^2 \mathbb{E} \{ |\vect{v}_{00k}^{\Htran} \vect{h}_{00k} |^2 \} +
 \sum_{i \neq k} p_{0i} \mathbb{E} \{ |\vect{v}_{00k}^{\Htran} \vect{h}_{00i} |^2 \} \notag
\\  &+ \sum_{j \in \Psi_{\lambda}}
\sum_{i=1}^{K} p_{ji} \mathbb{E} \{ | \vect{v}_{00k}^{\Htran} \vect{h}_{0ji} |^2 \}
+ \mathbb{E} \{ |\vect{v}_{00k}^{\Htran} \vect{n}_{0}|^2 \}  \Big) \label{eq:SINR-derivation}
\end{align}
where we have used the fact that $\mathbb{E}\{ | \sqrt{1-\epsilon^2}s_{ji} + e_{ji}  |^2  \} = 1-\epsilon^2 + \epsilon^2 = 1$. To proceed further with the computation of the expectations in \eqref{eq:SINR-derivation}, we define
 \begin{equation}
 \tau_{0k} = 1   + \sum_{l \in \Psi_{\lambda} }  \chi_{0kl}  \frac{d_{llk}^{\alpha}}{ d_{0lk}^{\alpha} } + \frac{\sigma^2}{ \rho}
 \end{equation}
and notice that the MMSE estimate of $\vect{h}_{0jk}$ (for $j=0$ or $j  \in \Psi_{\lambda}$ such that $\chi_{0kj} = 1$) takes the form 
 \begin{equation} \label{eq:MMSE-estimator-impairments}
 \begin{split}
  \hat{\vect{h}}_{0jk} &= 
\frac{  1}{ \tau_{0k} d_{0jk}^{\alpha} } \sqrt{ \frac{(1-\epsilon^2) d_{jjk}^{\alpha}}{\rho \omega } }\vect{y}_{0k}^{\textrm{pilot}} \\ &\sim \mathcal{CN}\left( \vect{0},  \frac{ (1-\epsilon^2)  d_{jjk}^{\alpha}   }{ \tau_{0k} \omega d_{0jk}^{2\alpha}} \vect{I}_M \right)
\end{split}
\end{equation}
whereas the independent estimation error is distributed as
\begin{equation}
\Delta \vect{h}_{0jk} \sim \mathcal{CN}\left( \vect{0}, \frac{1}{\omega d_{0jk}^{\alpha}} \left( 1 -  \frac{ (1-\epsilon^2) d_{jjk}^{\alpha}    }{ \tau_{0k} d_{0jk}^{\alpha}} \right) \vect{I}_M \right).
\end{equation}
For normalization purposes, the scaling factor $\nu_{00k}$ is selected as 
\begin{equation} \label{eq:nu-expression}
\nu_{00k} = \tau_{0k} \sqrt{\frac{  \omega d_{00k}^{\alpha}  }  { (1-\epsilon^2) \rho M  }}.
\end{equation}
Therefore, we have that
\begin{equation} \label{eq:express1}
\begin{split}
& p_{0k} |\mathbb{E} \{ \vect{v}_{00k}^{\Htran} \vect{h}_{00k} \}|^2  \\ &=\rho \omega d_{00k}^{\alpha} 
\left| \sqrt{\frac{  \omega d_{00k}^{\alpha}  }  { (1-\epsilon^2) \rho M  }} \mathbb{E} \{ \tau_{0k}  \hat{\vect{h}}_{00k}^{\Htran} \hat{\vect{h}}_{00k} \} \right|^2 \\ & =  (1-\epsilon^2)   M,
\end{split}
\end{equation}
which follows by using \eqref{eq:nu-expression} and the orthogonality principle of MMSE estimators as well as the adopted power-control policy in \eqref{p_{ji}}. Since the noise is independent of the channels, we also have 
\begin{equation} \label{eq:express2}
\begin{split}
 \mathbb{E} \{ | \vect{v}_{00k}^{\Htran} \vect{n}_{0}|^2 \} &= \sigma^2  \mathbb{E} \{ \nu_{00k}^2 \|\hat{\vect{h}}_{00k}\|^2 \} =  
  \frac{  \sigma^2 }{ \rho } \mathbb{E} \{ \tau_{0k} \}  \\ &=   \frac{  \sigma^2 }{ \rho }  \left( 1   + \frac{1}{\beta} \sum_{l \in \Psi_{\lambda} }  \frac{d_{llk}^{\alpha}}{ d_{0lk}^{\alpha} } + \frac{\sigma^2}{ \rho} \right).
  \end{split}
\end{equation}
Next, we consider the terms in \eqref{eq:SINR-derivation} that are affected by pilot contamination. For convenience, we let $\chi_{0k0}=1$ since the typical UE uses the same pilot as itself. Then, for all users $i$ in cells $j$ that are not causing pilot contamination (i.e., for $i \neq k$ or for $i=k$ and $\chi_{0kj}\ne1$), we have
\begin{equation} \label{eq:no-pilot-cont}
\begin{split}
& p_{ji} \mathbb{E} \{ | \vect{v}_{00k}^{\Htran} \vect{h}_{0ji} |^2 \} =
\frac{\rho \omega d_{jji}^{\alpha} }{ \omega d_{0ji}^{\alpha} } \mathbb{E} \{ \nu_{00k}^2 \| \hat{\vect{h}}_{00k}  \|^2 \} \\ &= 
\frac{d_{jji}^{\alpha} }{d_{0ji}^{\alpha}}
 \mathbb{E} \{ \tau_{0k} \} = \frac{d_{jji}^{\alpha} }{d_{0ji}^{\alpha}} \left( 1   + \frac{1}{\beta} \sum_{l \in \Psi_{\lambda} }  \frac{d_{llk}^{\alpha}}{ d_{0lk}^{\alpha} } + \frac{\sigma^2}{ \rho} \right)
 \end{split}
\end{equation}
where we have taken into account that $\hat{\vect{h}}_{00k}$ is independent of $\vect{h}_{0ji}$ and the last equality follows from averaging over the stochastic pilot selection (with $ \mathbb{E} \{ \chi_{0kl} \} = 1/\beta$ by design).

For all users $i$ in cells $j$ such that $i=k$ and $\chi_{0kj}=1$, the channel estimate $\hat{\vect{h}}_{00k}$ is correlated with $\vect{h}_{0ji}$ due to pilot contamination
and thus we obtain
\begin{align} \label{eq:with-pilot-cont}
& p_{jk} \mathbb{E} \{ | \vect{v}_{00k}^{\Htran} \vect{h}_{0jk} |^2 \} \\ &= \rho \omega d_{jjk}^{\alpha}  \Bigg( \frac{d_{jjk}^{\alpha}  d_{00k}^{\alpha} }{d_{0jk}^{2\alpha} } \mathbb{E} \{ \nu_{00k}^2 | \hat{\vect{h}}_{00k}^{\Htran} \hat{\vect{h}}_{00k} |^2 \} \notag \\ & \qquad + \mathbb{E} \{ \nu_{00k}^2 | \hat{\vect{h}}_{00k}^{\Htran} \Delta \vect{h}_{0jk} |^2 \}  \Bigg) \notag\\
&=  \omega d_{jjk}^{\alpha}  \Bigg[ \frac{d_{jjk}^{\alpha}  d_{00k}^{\alpha} }{d_{0jk}^{2\alpha} }
 \frac{ (1-\epsilon^2 )  }{   \omega d_{00k}^{\alpha} }  (1+M) \notag \\ & \qquad + 
\frac{1}{\omega d_{0jk}^{\alpha}} \left(  \mathbb{E} \{ \tau_{0k} \} -  \frac{ (1-\epsilon^2) d_{jjk}^{\alpha}    }{ d_{0jk}^{\alpha}} \right) \Bigg] \notag \\
 &=  \left( \frac{d_{jjk}^{\alpha} }{d_{0jk}^{\alpha} } \right)^2 \! \! (1-\epsilon^2) M
  +  \frac{d_{jjk}^{\alpha} }{d_{0jk}^{\alpha} }  \! \left( 1   + \frac{1}{\beta} \sum_{l \in \Psi_{\lambda} }  \frac{d_{llk}^{\alpha}}{ d_{0lk}^{\alpha} } + \frac{\sigma^2}{ \rho} \right)
\end{align}
where the first equality follows from the expansion $\vect{h}_{0jk} = \hat{\vect{h}}_{0jk} + \Delta \vect{h}_{0jk} $ and the fact that $\hat{\vect{h}}_{0jk} = \sqrt{\frac{d_{jjk}^{\alpha}  d_{00k}^{\alpha} }{d_{0jk}^{2\alpha} }
} \hat{\vect{h}}_{00k}$. The second equality follows from using \cite[Lemma 2]{Bjornson2015b} and \eqref{eq:nu-expression}, whereas the third equality simplifies the expression taking into account that  $ \mathbb{E} \{ \chi_{0kl} \} = 1/\beta$. Since the second term in \eqref{eq:with-pilot-cont} is equal to \eqref{eq:no-pilot-cont}, it follows that it is only the first term in \eqref{eq:with-pilot-cont} that comes from pilot contamination and its probability to appear is $ \mathbb{E} \{ \chi_{0kl} \} = 1/\beta$. Plugging \eqref{eq:express1}--\eqref{eq:with-pilot-cont} into \eqref{eq:SINR-derivation} the result in the lemma follows.

\section*{Appendix B: Proof of Proposition \ref{prop:average-SE}}
\label{app:prop:average-SE}

The goal is to compute a tractable lower bound on the average SE $( 1 - \frac{B}{S} ) \mathbb{E}\{ \log_2 ( 1 + \mathrm{SINR}_{0k}  ) \}$, where the expectation is taken with respect to the PPP and UE locations.
To this end, the Jensen's inequality is applied as
$\mathbb{E}\{ \log_2(1+\frac{1}{\mathrm{SINR}_{0k}^{-1} }) \} \geq  \log_2(1+\frac{1}{\mathbb{E}\{ \mathrm{SINR}_{0k}^{-1} \}})$ to move the expectation inside the logarithm. The expectation of the inverse SINR is then expanded as in \eqref{eq:inverse-SINR} at the top of the page. \setcounter{equation}{51}

To proceed further, we focus on the BSs in a circular area of finite radius $r$ and consider wrap around in the radial domain to keep the translation invariance. Note that by letting $r \rightarrow \infty$ a PPP in the whole of $\mathbb{R}^2$ is obtained, and there is no bias since the intra-cell distances are assumed to be distributed exactly as in $\mathbb{R}^2$. The reason for doing this is that the average number of interfering BSs in the limited circular area is finite and given by $\pi (r^2 - \mathbb{E}\{ d_{00k}^2 \} ) = \pi (r^2 - \frac{1}{\pi \lambda} )$, where we have used  the fact that $d_{00k} \sim \mathrm{Rayleigh} \big( \frac{1}{\sqrt{2\pi \lambda}} \big)$
 (see Lemma \ref{lemma:distance-distribution}) to obtain
\begin{equation} \label{eq:dook_distance}
\mathbb{E}\{ d_{00k}^\nu \} =  \frac{\Gamma(\nu/2+1)}{(\pi \lambda)^{\nu/2}}
\end{equation}
for any $\nu > - 2$. 
Two of the expectations in \eqref{eq:inverse-SINR} contain summations over the PPP $\Psi_{\lambda}$, but with different powers of the terms. To compute both, we let $\kappa=1$ or $\kappa=2$ and notice that
\begin{align} \notag
&\mathbb{E} \left\{ \fracSum{j \in \Psi_{\lambda} }  
\left( \frac{ d_{jjk}^{\alpha} }{ d_{0jk}^{\alpha}} \right)^{\kappa} \right\}  \stackrel{(a)}{=} \lambda \pi \left(r^2 - \frac{1}{\pi \lambda} \right)  \mathbb{E} \left\{ \left( \frac{ d_{jjk}^{\alpha} }{ d_{0jk}^{\alpha}} \right)^{\kappa} \right\}  \\ & \notag \stackrel{(b)}{=} \lambda \pi \left(r^2 - \frac{1}{\pi \lambda} \right)   \mathbb{E}\left\{  \mathbb{E}\left\{ 
d_{0jk}^{-\kappa \alpha} 
  \Big| d_{jjk} \right\} d_{jjk}^{\kappa \alpha} \right\} \\ \notag
&\stackrel{(c)}{=} \lambda \pi \left(r^2 - \frac{1}{\pi \lambda} \right)  \mathbb{E}\left\{ 
d_{jjk}^{\kappa \alpha} \int_{d_{jjk}}^{r} x^{-\kappa \alpha} \frac{2x}{r^2-d_{jjk}^2} dx  \right\} \\ &
\stackrel{(d)}{=} \lambda \pi \left(r^2 - \frac{1}{\pi \lambda} \right) \mathbb{E}\left\{ d_{jjk}^{\kappa \alpha} \frac{2}{r^2-d_{jjk}^2} 
\frac{d_{jjk}^{2-\kappa \alpha} - r^{2-\kappa \alpha}}{\kappa \alpha-2} \right\} \notag \\  
&  
\xrightarrow{(e): \, r \to \infty} \frac{2 \pi \lambda}{\kappa \alpha-2 } \mathbb{E}\left\{ 
d_{jjk}^{2} \right\} = \frac{2 \pi \lambda}{\kappa \alpha-2 } \frac{\Gamma(2)}{\pi \lambda} = \frac{2 }{\kappa \alpha-2} \label{eq:derivation-moment}
\end{align}
where $(a)$ follows from computing the average number of BSs in the circular area and by letting $j$ be an arbitrary BS index in $\Psi_{\lambda}$ and $k$ be an arbitrary UE index (since each term in the sum has the same marginal distribution). Next, $(b)$ divides the expectation into one conditional expectation where $d_{jjk}$ is given and one outer expectation with respect to $d_{jjk}$. Step $(c)$ and $(d)$ compute the inner expectation by utilizing that the BSs of other cells are uniformly distributed in the circle of radius $r$ at distances larger than $d_{jjk}$. The limit $r \rightarrow \infty$ in $(e)$ is taken both inside and outside the expectation, which is allowed since the dominated convergence theorem is satisfied.\footnote{The dominated convergence theorem can be applied since $|d_{jjk}^{\alpha} /d_{0jk}^{\alpha}| \leq 1$.} The final expression is obtained by computing $\mathbb{E}\{ 
d_{jjk}^{2} \}$ as in \eqref{eq:dook_distance} and exploiting the typicality, which implies that $d_{jjk}$ and $d_{00k}$ have the same marginal distribution. Using similar techniques as in \eqref{eq:derivation-moment} yields
\begin{equation}
\mathbb{E} \left\{ \fracSum{j \in \Psi_{\lambda}} \fracSumtwo{i=1}{K} 
\frac{d_{jji}^{\alpha}}{ d_{0ji}^{\alpha} } \right\} = K \frac{ 2}{\alpha-2}.
\end{equation}
The only remaining term in \eqref{eq:inverse-SINR} is
\begin{equation} \label{eq:second-moment}
\begin{split}
& \mathbb{E} \left\{ \left(  \fracSum{j \in \Psi_{\lambda}} \fracSumtwo{i=1}{K} 
\frac{d_{jji}^{\alpha}}{ d_{0ji}^{\alpha} }   \right) \left(  \fracSum{l \in \Psi_{\lambda} }   \frac{d_{llk}^{\alpha}}{ d_{0lk}^{\alpha} } \right) \right\} \\ &= \fracSumtwo{i=1}{K} 
\mathbb{E} \left\{ \fracSum{j \in \Psi_{\lambda} }
\fracSum{l \in \Psi_{\lambda} \setminus \{ j \} } \frac{d_{jji}^{\alpha}}{ d_{0ji}^{\alpha} } \frac{d_{llk}^{\alpha}}{ d_{0lk}^{\alpha} } \right\} 
\\ &+ \fracSumtwo{i=1}{K} \mathbb{E} \left\{ \fracSum{l \in \Psi_{\lambda} } \frac{d_{lli}^{\alpha}}{ d_{0li}^{\alpha} } \frac{d_{llk}^{\alpha}}{ d_{0lk}^{\alpha} } 
\right\} 
\end{split}
\end{equation}
where the equality follows from computing the expectations separately over sets that contain different BS indices.
The first term in \eqref{eq:second-moment} is computed as \\
\begin{equation}\label{1000}
\begin{split}
&\fracSumtwo{i=1}{K} 
\mathbb{E} \left\{ \fracSum{j \in \Psi_{\lambda} }
\fracSum{l \in \Psi_{\lambda} \setminus \{ j \} } \frac{d_{jji}^{\alpha}}{ d_{0ji}^{\alpha} } \frac{d_{llk}^{\alpha}}{ d_{0lk}^{\alpha} } \right\} \\ & = K 
\left( \lambda \pi \left(r^2 - \frac{1}{\pi \lambda} \right) \right) \left( \lambda \pi \left(r^2 - \frac{1}{\pi \lambda} \right) -1 \right)  \\ & \quad \times
\mathbb{E}\left\{ d_{jji}^{\kappa \alpha} \frac{2}{r^2-d_{jji}^2} 
\frac{d_{jji}^{2-\kappa \alpha} - r^{2-\kappa \alpha}}{\kappa \alpha-2} \right\} \\ & \quad \times
\mathbb{E}\left\{ d_{llk}^{\kappa \alpha} \frac{2}{r^2-d_{llk}^2} 
\frac{d_{llk}^{2-\kappa \alpha} - r^{2-\kappa \alpha}}{\kappa \alpha-2} \right\}  \\ 
& \xrightarrow{r \to \infty}  
 K \pi^2 \lambda^2 \! \left( \frac{2}{\alpha-2}\right)^2 \mathbb{E}\left\{ d_{jji}^{2}  \right\} 
\mathbb{E}\left\{ d_{llk}^{2} \right\} \\& = K \pi^2 \lambda^2 \! \left( \frac{2}{\alpha-2}\right)^2 \! \left(\frac{\Gamma(2)}{\pi \lambda}\right)^2 \!= K \! \left( \frac{2}{\alpha-2}\right)^2
\end{split}
\end{equation}
by first computing the expectation with respect to the number of terms in the PPP for a finite radius $r$ and exploiting that $j \neq l$. By letting $r \rightarrow \infty$ we arrived at an expression that is basically twice the expression previously given in \eqref{eq:derivation-moment}.
The second term in \eqref{eq:second-moment} is bounded as
\begin{align} \notag
 & \fracSumtwo{i=1}{K} \mathbb{E} \left\{ \fracSum{l \in \Psi_{\lambda} } \frac{d_{lli}^{\alpha}}{ d_{0li}^{\alpha} } \frac{d_{llk}^{\alpha}}{ d_{0lk}^{\alpha} } \right\} \! \stackrel{(a)}{=} \! K \lambda \pi \left(r^2 - \frac{1}{\pi \lambda} \right)  \notag
 \mathbb{E} \left\{ \frac{d_{lli}^{\alpha}}{ d_{0li}^{\alpha} } \frac{d_{llk}^{\alpha}}{ d_{0lk}^{\alpha} } \right\} \\ &\notag \stackrel{(b)}{\leq} K \lambda \pi \left(r^2 - \frac{1}{\pi \lambda} \right)  \sqrt{
 \mathbb{E} \left\{ \frac{d_{lli}^{2\alpha}}{ d_{0li}^{2\alpha} } \right\}   \mathbb{E} \left\{ \frac{d_{llk}^{2\alpha}}{ d_{0lk}^{2\alpha} } \right\} } \\ & \stackrel{(c)}{=}  K \lambda \pi \left(r^2 - \frac{1}{\pi \lambda} \right) \mathbb{E} \left\{ \frac{d_{llk}^{2\alpha}}{ d_{0lk}^{2\alpha} } \right\} \notag \\ & \xrightarrow{(d): \, r \to \infty} 
K \frac{2 }{2 \alpha-2 } = K \frac{1 }{ \alpha-1 } \label{1001}
\end{align}
where $(a)$ computes the expectation with respect to the number of terms in the PPP for a finite $r$, $(b)$
follows from H\"{o}lder's inequality, and $(c)$ exploits that the two expectations are equal. 
At this point, the expression is basically the same as the second expression in \eqref{eq:derivation-moment}, thus $(d)$ follows from the same steps as those taken in \eqref{eq:derivation-moment}. Plugging \eqref{1000} and \eqref{1001} into \eqref{eq:second-moment} we eventually obtain 
\begin{equation} \label{eq:second-moment2}
\begin{split}
& \mathbb{E} \left\{ \left(  \fracSum{j \in \Psi_{\lambda}} \fracSumtwo{i=1}{K} 
\frac{d_{jji}^{\alpha}}{ d_{0ji}^{\alpha} }   \right) \left(  \fracSum{l \in \Psi_{\lambda} }   \frac{d_{llk}^{\alpha}}{ d_{0lk}^{\alpha} } \right) \right\} \\ & \leq  K \left( \frac{4}{(\alpha-2)^2 } +
\frac{1 }{ \alpha-1} \right).
\end{split}
\end{equation}
We have now computed all the expectations in \eqref{eq:inverse-SINR}, exactly or as upper bounds. Using all these expressions eventually lead to the achievable lower bound in \eqref{eq:average-SINR}.

\bibliographystyle{IEEEbib}
\bibliography{IEEEabrv,refs}

\end{document}